\documentclass[aps, pre, reprint, superscriptaddress, longbibliography, showkeys, nofootinbib, floatfix, nobalancelastpage ]{revtex4-2}

\usepackage{amssymb, mathtools, amsmath, amsfonts, physics, blkarray, booktabs}
\usepackage{graphicx, color, soul, cancel, lipsum, xspace, tikz, placeins, braket, fontawesome}
\usepackage[colorlinks=true,linkcolor=blue,urlcolor=blue,citecolor=blue,pdfusetitle]{hyperref}

\usepackage[utf8]{inputenc}
\usepackage[T1]{fontenc}

\usepackage{graphicx}
\usepackage{dcolumn}
\usepackage{bm}

\usepackage{amsmath,amsthm,amsfonts,amssymb,bbm,mathtools, xcolor}
\newcommand{\matr}[1]{\mathbf{#1}}

\newsavebox{\mstrut}
\newcommand{\bbra}[1]{%
    \sbox{\mstrut}{\(#1\)}%
    \mathinner{\left\langle\kern-0.5\ht\mstrut\left\langle{#1}\right|\mkern-2mu\right|}%
}
\newcommand{\kket}[1]{%
    \sbox{\mstrut}{\(#1\)}%
    \mathinner{\left|\mkern-2mu\left|{#1}\right\rangle\kern-0.5\ht\mstrut\right\rangle}%
}

\begin{document}

\preprint{} 

\title{Optimal switching strategies in multi-drug therapies for chronic diseases}
\author{Juan Magalang}
\affiliation{%
Department of Visceral Surgery and Medicine, Inselspital, Bern University Hospital, University of Bern,
Murtenstrasse 35, Bern, 3008, Switzerland
}%
\affiliation{Institute of Mathematical Statistics and Actuarial Science, University of Bern, Alpeneggstrasse 22, Bern, 3012, Switzerland
}%

\author{Javier Aguilar}%
\affiliation{Laboratory of Interdisciplinary Physics, Department of Physics and Astronomy
“G. Galilei”, University of Padova, Padova, Italy}%

\author{Jose Perico Esguerra}
\affiliation{Theoretical Physics Group, National Institute of Physics, University of the Philippines Diliman, 
Quezon City, 1101, Philippines
}%

\author{\'Edgar Rold\'an}
\email{edgar@ictp.it}
\affiliation{%
ICTP - The Abdus Salam International Centre for Theoretical Physics,
Strada Costiera 11, Trieste, 34151, Italy
}%

\author{Daniel Sanchez-Taltavull}
\email{daniel.sanchez@unibe.ch}
\affiliation{%
Department of Visceral Surgery and Medicine, Inselspital, Bern University Hospital, University of Bern,
Murtenstrasse 35, Bern, 3008, Switzerland
}%

\date{\today}

\begin{abstract}
Antimicrobial resistance is a threat to public health with millions of deaths linked to drug resistant infections every year. To mitigate resistance, common strategies that are used are combination therapies and therapy switching. However, the stochastic nature of pathogenic mutation makes the optimization of these strategies challenging. Here, we propose a two-scale stochastic model that considers the effective evolution of therapies in a multidimensional efficacy space, where each dimension represents the efficacy of a specific drug in the therapy. The diffusion of therapies within this space is subject to stochastic resets, representing therapy switches. The boundaries of the space, inferred from coarser pathogen-host dynamics, can be either reflecting or absorbing. Reflecting boundaries impede full recovery of the host, while absorbing boundaries represent the development of antimicrobial resistance, leading to therapy failure. We derive analytical expressions for the average absorption times, accounting for both continuous and discrete genomic changes using the frameworks of Langevin and Master equations, respectively. These expressions allow us to evaluate the relevance of times between drug-switches and the number of simultaneous drugs in relation to typical timescales for drug resistance development. We also explore  realistic scenarios where therapy constraints are imposed to the number of administered therapies and/or their costs, finding non-trivial optimal drug-switching protocols that maximize the time before antimicrobial resistance develops while reducing therapy costs.

\end{abstract}

\maketitle

\section{\label{sec:intro} Introduction}

Antimicrobial resistance has been recognized as a major threat to public health, with estimates of 4 million deaths associated with resistant bacterial infection in 2019~\cite{Murray2022}. This resistance is a consequence of pathogenic evolution, affecting multiple diseases such as tuberculosis~\cite{liebenberg2022drug}, HIV/AIDS~\cite{phillips2017impact, Drechsler2002}, cancer~\cite{vasan2019view,catalano2022multidrug,maltas2024}, among others. Drug resistant pathogens are more difficult to treat since conventional treatment procedures can no longer be used. This increases the risk of complications due to more aggressive drugs and the economic burden, with excess costs estimated at one hundred thousand US dollars per case of multi-drug resistant tuberculosis in 2013 \cite{Naylor2018}. The extent and severity of drug resistance is expected to increase in the coming years~\cite{ONEILLAMR,langford2023antimicrobial}, creating the need to determine optimal treatment strategies to minimize the problem. 

A common strategy that physicians use to avoid drug resistance is by combining multiple drugs \cite{baym2016multidrug,tyers2019drug}. Combination therapies target multiple biological mechanisms of a pathogen, reducing the likelihood of developing resistance to all drugs combined. When resistances occur, physicians typically replace the therapy \cite{band2019antibiotic,Drechsler2002, Haas2015}. However, stochastic effects can play a role in either the detection of resistance \cite{Keiser2009, Reynolds2009}, the management of adverse effects \cite{mayer2002switching, DAVIDSON2010227}, or the introduction of new drugs in the market \cite{Kirby2019}. Therefore, to fully understand the problem, we need to simultaneously account for the stochastic effects in therapy switching and the multiple components of combination therapy.

Mathematical models have been used to emulate drug resistance at different scales, from the cellular mechanisms \cite{Pinheiro2021, Andersson2019}, to within-host dynamics \cite{Lee2022, Rong2009, Callaway2002, Sharomi2008}, and the overall epidemiological and economic impact \cite{Niewiadomska2019, Heesterbeek2015, Hillock2022}. These models allowed for the study of the effectiveness of public health strategies that mitigate resistance, such as improving hygiene protocols, increasing surveillance, and regulating the use of antimicrobial drugs \cite{Niewiadomska2019}.

Our goal is to develop a stochastic model that can identify optimal strategies of therapy administration and therapy switching rates. In doing so, we present a two-scale model that describes the efficacy of therapies administered to a host. The first scale accounts for pathogenic evolution which results in changes in the therapy efficacy. This evolution is a direct consequence of pathogenic mutation which is inherently stochastic~\cite{Manrubia2012, papkou2023rugged}. The second scale accounts for host-pathogen dynamics of a chronic infection, such as HIV-1 \cite{Rong2009, Callaway2002} and is linked to the first scale via the infection rate which depends on the current therapy efficacy. 

Pathogenic evolution is modeled as a diffusion process in multiple dimensions, where each component of the position vector of the process represents the efficacy of a drug, and therapy changes are modeled using the framework of stochastic resetting~\cite{evans_diffusion_2011, evans_stochastic_2020}, which has found recent applications in biology~\cite{rotbart2015michaelis,roldan2016stochastic,lisica2016mechanisms,bressloff2020modeling}.  While a faithful representation of evolutionary dynamics should bear the discreteness of genomic changes~\cite{papkou2023rugged}, continuous approximations of these dynamics are also used~\cite{wang2011quantifying} to ease the analysis. The efficacy space also includes a region representing therapy failure. Hence, drug resistance development becomes a first passage time problem \cite{redner2001guide}. Typically, stochastic resetting problems aim to minimize the first passage time \cite{Evans2011, Pal2019, Kumierz2015, Pal2017, Das2022}, but the need to maximize this time emerges naturally, as it is equivalent to the drug resistance development time \cite{ramoso_stochastic_2020}. 

The paper is organized as follows: We first discuss the specifications of the model (Section \ref{sec:models}). Then we analyze the dependence of the model on its parameters (Section \ref{sec:results1}) and the effect of imposing restrictions to the number of administered therapies (Section \ref{sec:results2}). Finally, we summarize and discuss our results (Section \ref{sec:discussion}).

\section{\label{sec:models}  Multi-drug Stochastic models}

We are interested in modelling a host infected with a pathogen that accounts for the development of drug resistance in two time scales: pathogenic evolution and the host-pathogen dynamics. These two scales are linked by an infection rate, which determines the number of infected cells in the host-pathogen model, and it is reduced by the therapy efficacy, similar to previous models \cite{Rong2009,Callaway2002, Sharomi2008}. It is known that pathogenic evolution is influenced by more stochastic effects and has slower dynamics than the host-pathogen interactions \cite{Manrubia2012, papkou2023rugged}. Hence, we choose to focus on the scale of evolution and the host-pathogen dynamics are assumed to be at the steady state. By modelling the fluctuation of therapy efficacy due to evolution, we aim to obtain approximations to the time at which drug resistance occurs, which we define as the \textit{resistance development time} (RDT). This quantity is synonymous with the first passage time, defined as the time at which a stochastic process reaches a certain state \cite{redner2001guide}.

\begin{figure*}[!htb]
    \centering \includegraphics[width=\textwidth]{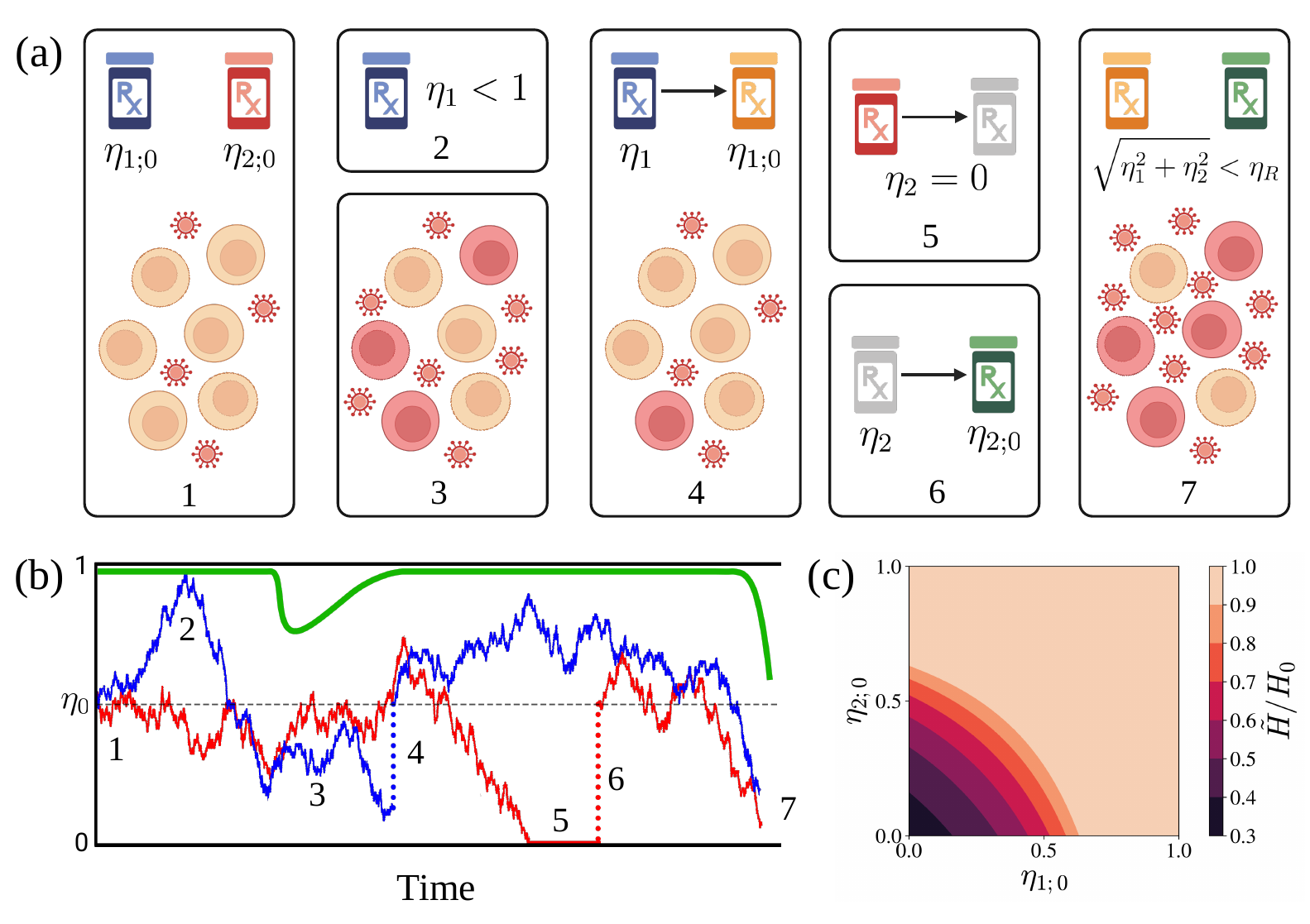}
    \caption{Schematic representation of the two-scale mathematical model of drug resistance development in terms of the efficacy of an antimicrobial therapy. (a) A diagram showing the host-pathogen scale for two therapies, illustrating how drug resistance emerges and the impact of therapy changes. (b) Illustration of a stochastic trajectory of the efficacies of two therapies (in blue and red) and the normalized number of healthy cells ($H/H_0$, in green), showing the events described in panel (a): (1) The initial value of therapy efficacy is shown as a black horizontal dashed line. (2) The therapy efficacy is bounded by a reflecting boundary at $\eta = 1$. (3) Healthy cells drop due to a decrease in efficacy of both therapies. (4) Switching one of the therapies. (5) \textit{Partial} absorption due to one of the therapies failing, and does not decrease the healthy population due to the effect of the other therapy. (6) Switching the failed therapy. (7) Failure of both therapies leading to a decrease in the healthy cells to a critical level. (c) Normalized number of healthy cells at equilibrium as a function of the therapy efficacies, showing the existence of a critical region when both efficacies are low. The analytical expression for the equilibrium of $H$ is computed in Appendix \ref{AP:fixedpt} and within-host model parameters are found in Table \ref{table:params}.
}
    \label{fig:2thdiagram}
\end{figure*}

Our model considers $N_T\geq 1$ simultaneous and independent therapies, of which the $i$\textsuperscript{th} therapy efficacy $\eta_i$ undergoes evolution as a stochastic process. The therapy efficacy scales an infection rate given by $\prod_{i=1}^{N_T}(1-\eta_i)\beta H I$, where $\beta$ is the infection rate constant, and $H$ is the population of healthy, susceptible cells, and $I$ is the population of actively-infecting cells. The form of the infection rate implies that therapy efficacy is bounded: $\eta_i \in [0,1) \, \forall i$. A therapy efficacy of $\eta_i = 1$ signifies a perfect therapy, as it removes the infection rate. On the other hand, $\eta_i = 0$ refers to an absorbing boundary and signifies a complete failure of the therapy as it maximizes the rate. Figure~\ref{fig:2thdiagram} illustrates the stochastic process being used to model how therapy efficacy changes over time. Further details of the host-pathogen model are provided in Appendix~\ref{AP:fixedpt}. 

Note that it is not necessary for all $N_T$ therapies to reach complete failure for the host to be in a critical state due to an infection of a drug-resistant pathogen. In Fig. \hyperref[fig:2thdiagram]{1c} we begin to see a decrease in the number of healthy cells even with $\eta_i > 0$. Hence, we consider that the therapy efficacy is below a certain threshold for drug resistance to occur. In this work, we define $\sqrt{\sum_{i=1}^{N_T} \eta_i^2} \leq \eta_{\text{min}}$ as the condition for drug failure, with $\eta_{\text{min}} = 0.4$.

An essential ingredient of our model is the inclusion of therapy switching at stochastic times. When a therapy is switched with a different one, the pathogens are exposed to a new stimulus and must restart the drug resistance development process; we model such switches as a stochastic resetting process~\cite{evans_diffusion_2011, evans_stochastic_2020}. See the illustration of therapy switches in  events 4 and 5 of Figs. ~\hyperref[fig:2thdiagram]{1a} and ~\hyperref[fig:2thdiagram]{1b}. Since we describe the evolution of the multi-therapy efficacy as a stochastic process, we may use the following multivariate stochastic differential equation (SDE)
\begin{equation}\label{eq:SDE_general}
\begin{aligned}
    \mathrm{d} \pmb{\eta} =& \, \mathrm{diag}\left(\pmb{1}-\pmb{\mathcal{I}}_{\pmb{\chi}}(t) \right) \left[ \pmb{\mu}(\pmb{\eta}, t) \mathrm{d}t + \sqrt{2D} \, \mathrm{d} \pmb{W}(t) \right] \\
    &+ \mathrm{diag} \left(\pmb{\mathcal{I}}_{\pmb{\chi}}(t) \right) (\pmb{\eta}_0 - \pmb{\eta}),
\end{aligned}
\end{equation}
where $\pmb{\eta} = \{\eta_1, \eta_2, \ldots, \eta_{N_T}\}$ is a vector containing the $N_T$ therapy efficacies, $\pmb{\mu}(\pmb{\eta},t)$ is a drift vector affecting each of the therapies, $D$ is the diffusivity  (assumed to be isotropic), and $\pmb{W}(t)$ a $N_T-$dimensional Wiener process. Pathogenic evolution favors mutations with stronger resistances to improve the survival~\cite{Manrubia2012}, therefore we assume that the drift vector is biased towards therapy failure, $\pmb{\mu}_i<0$. The operator $\mathrm{diag}(\cdot)$ transforms a vector argument into a square diagonal matrix. 

The quantity $\pmb{\mathcal{I}}_{\pmb{\chi}}(t)$ is a vector of indicator functions associated to $\pmb{\chi}$, which controls the return of the therapy efficacies to its corresponding initial value $\pmb{\eta}_0$. This vector has elements $    \mathcal{I}_{\chi_i}(t) =1$ for  $t \in \chi_i$, and $    \mathcal{I}_{\chi_i}(t) =0$ for  $t \notin \chi_i$, 
where $\chi_i = \{0, t^*_{i,1}, t^*_{i,2}, \ldots \}$ for all $1\leq i \leq N_T$. Here,~$\chi_i$ is a sequence of partial sums of i.i.d. exponentially-distributed random numbers, i.e. $t^*_{i,j}-t^*_{i,j-1} \sim \text{Exp}(\tau_i)$, with mean $1/\tau_i$, which is the therapy switching rate for the $i$\textsuperscript{th} therapy.

Solving the RDT statistics associated with the SDE~\eqref{eq:SDE_general} is challenging due to the geometry of the boundary conditions $\eta_{\min}$ and standard techniques to solve this equation are not known to be effective \cite{redner2001guide, dyesguerra, Gardiner2008-re, giuggioli2023}. We instead propose two models of an $N_T$-dimensional fluctuating therapy efficacy with therapy switching that are analogous to Eq.~\eqref{eq:SDE_general} that exploits symmetries and approximations in the space spanned by the therapy efficacy. The first method is a continuous-space model that reduces the dynamics of the multidimensional SDE which allows for the calculation of analytical expressions of the mean RDT. The second method is a discrete-space model that explicitly shows the dynamics of the model and produces all statistical quantities of the RDT.

\begin{figure*}[!htb]
    \centering
    \includegraphics[width=0.85\textwidth]{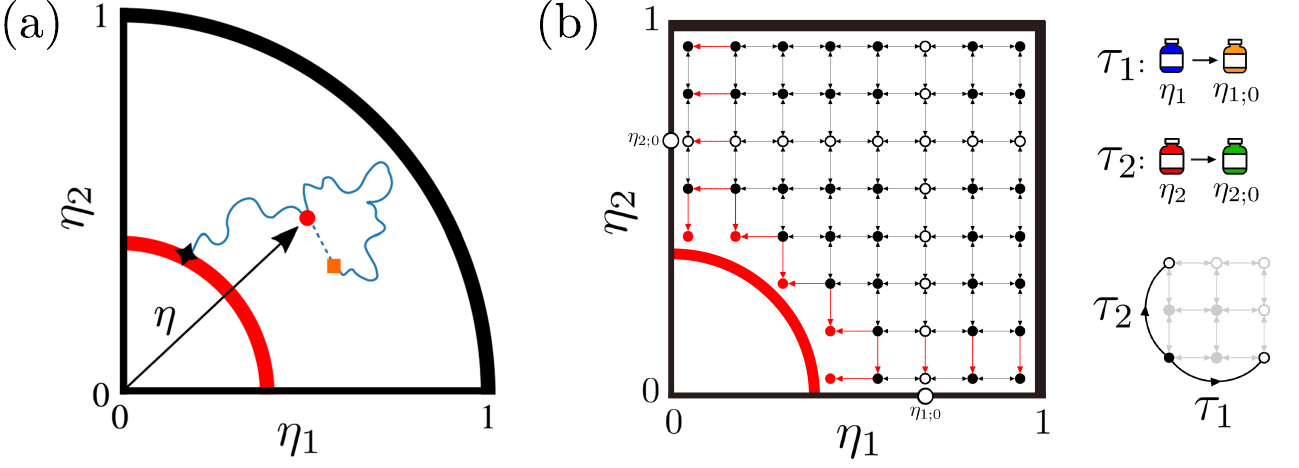}
    \caption{Sketch of the coupled and uncoupled models for $N_T = 2$ for the efficacy of two drugs.  (a) The coupled continuous model, with rotational symmetry and $\eta$ as the distance from the origin. The trajectories start at the initial efficacy (dot), upon a therapy switch (square) the therapy efficacy goes back to the initial position, and the process stops when it reaches the absorbing region (star). (b) The uncoupled discrete model forms a lattice of $M \times M$ states, where each state corresponding to a value of the efficacy. Changes in therapy efficacy are transitions from one state to an adjacent state. Red states represent the absorbing states of the model. Upon a therapy switch in $i$\textsuperscript{th} therapy, the system returns to the points labelled in white, $\eta_{i;0}$. Efficacy $\eta_i = 1$ represent reflecting boundaries and $\eta_i =0$ represents partially absorbing boundaries.
}
    \label{fig:multi-drug_model}
\end{figure*}

\subsection{Coupled Continuous Model} \label{subsec:coupled}

In our first model of drug resistance development, we assume that the discrete changes in the genome of the pathogen are small enough to be considered continuous \cite{wang2011quantifying}. Furthermore, we consider that the behavior of the multi-drug therapy efficacy is expressed as a single parameter: $\eta=\sqrt{\sum_{i=1}^{N_T} \eta_i^2}$, motivated by the absorbing boundary described in Fig. \hyperref[fig:2thdiagram]{1c}, and evolves following the SDE:
\begin{equation}\label{eq:SDE_coupled}
    \mathrm{d}\eta = \left(1-\mathcal{I}_{\chi}(t) \right) \left[ \frac{\tilde{v}}{\eta} \mathrm{d}t + D \mathrm{d} W(t) \right] + \mathcal{I}_{\chi}(t) (\eta_0 - \eta),
\end{equation}
where $\tilde{v}=D [N_T-1- (v/D)]$ is an effective drift. Even when the process is one dimensional, the effective drift still depends explicitly on the number of dimensions $N_T$. Equation~\eqref{eq:SDE_coupled} also imposes a rotational symmetry in the efficacy space. This symmetry respects the absorbing boundary in Fig.~\hyperref[fig:2thdiagram]{1c}, while allowing us to study the efficacy evolution of a therapy with an arbitrary number of drugs. A schematic of this model is found in Fig. \hyperref[fig:multi-drug_model]{2a} for $N_T = 2$ and the detailed derivation of Eq.~\eqref{eq:SDE_coupled} is found in Appendix~\ref{AP:Coupled_model}. Using the formalism of the backward Fokker-Planck equation, we derive an analytical expression for the mean RDT conditioned to an initial efficacy $\eta$  \cite{vankampen2007,Gardiner2008-re},
\begin{widetext}
\begin{equation}\label{eq:AFPT_coupled}
   \langle T \vert \eta \rangle  =\tau \, \frac{Y_d(- \mathrm{i} \overline{\eta}_{\max})\left[J_{d-1}\left(- \mathrm{i} \overline{\eta} \right)-\left(\eta/\eta_\text{min}\right)^{d-1}J_{d-1}(- \mathrm{i} \overline{\eta}_{\min})\right]+J_d(- \mathrm{i} \overline{\eta}_{\max})\left[Y_{d-1}\left(- \mathrm{i}\overline{\eta}\right)-\left(\eta/\eta_\text{min}\right)^{d-1}Y_{d-1}(- \mathrm{i} \overline{\eta}_{\min})\right]}{J_d(- \mathrm{i}{\overline{\eta}_{\max}})Y_{d-1}(- \mathrm{i}\overline{\eta})-Y_d(-\mathrm{i} \overline{\eta}_{\max})J_{d-1}(- \mathrm{i}\overline{\eta})}.
\end{equation}
\end{widetext}
Here, $J_n$, $Y_n$ are the Bessel functions of order $n$ of the first and second kind respectively, and $d = 2[(v/D)+N_T]$, where $(v/D)+N_T$  can be interpreted as an effective dimension of the efficacy space. 
We also introduced rescaled efficacies $\overline{\eta} = \lambda \eta$,  $\overline{\eta}_{\min} = \lambda \eta_{\min} $, with $\lambda=(D\tau)^{-1/2}$ and $\overline{\eta}_{\max} = \lambda \eta_{\max}$, with $\eta_{\min}$ and $\eta_{\max}$ as the scaled radial locations of the absorbing and reflecting boundary respectively. The characteristic scale $\lambda$ is
the inverse of the typical length traveled by genetic diffusion before  a therapy switch occurs. See Appendix~\ref{AP:Coupled_model} for the derivation of Eq.~\eqref{eq:AFPT_coupled} and further mathematical details.

\subsection{Uncoupled Discrete Model}\label{subsec:uncoupled}

In the second model of drug resistance, we model the discrete phenotypic changes of a pathogen undergoing mutation \cite{papkou2023rugged} as a chain of states. We assume that the chain of states is equally-spaced with $M$ states, such that $\eta_i = j/M$, for $j = 0, \ldots, M-1$ and $i = 1, \ldots, N_T$. Transition rates control the evolution of $\eta_i$ on the chain: rate $p_i$ refers to an increase $\eta_i \to \eta_i + 1/M$, rate $q_i$ refers to a decrease $\eta_i \to \eta_i - 1/M$, and rate $1/\tau_i$ refers to a therapy switch $\eta_i \to \eta_{i;0}$. This model independently evolves and switches the therapy efficacy for all $N_T$ therapies.

For simultaneous therapies, we form a vector of therapy efficacies $\pmb{\eta} = (\eta_1, \ldots, \eta_{N_T})^\top$, which is interpreted as an ordered coordinate, generalizing the chain of $M$ states to an $N_T$-dimensional lattice of $M^{N_T} \times M^{N_T}$ states. A sketch of how this model is constructed is in Fig. \hyperref[fig:multi-drug_model]{2b} for $N_T = 2$. Transitions between states on this lattice can be written as a transition matrix $\matr{W}$, and the evolution of the vector $\pmb{\eta}$ is controlled by a master equation
\begin{equation}\label{eq:gridmastereqn}
    \frac{\mathrm{d} p(\pmb{\eta} ,t)}{\mathrm{d}t} = \matr{W} p(\pmb{\eta},t),
\end{equation}
where $p(\pmb{\eta}, t)$ is a column vector of probabilities tracking all $N_T$ therapy efficacies at each time. For any two efficacy states on the lattice $\pmb{\eta}_a$ and $\pmb{\eta}_b$, $\matr{W}$ has positive elements $W_{ba}$ referring to the transition rates from state $\pmb{\eta}_a$ to $\pmb{\eta}_b$, and negative diagonal elements $W_{bb} = - \sum_{a \neq b} W_{ba}$ referring to the escape rates from state $\pmb{\eta}_b$. Further details about the construction of the transition matrix $\matr{W}$ are found in Appendix \ref{AP:uncoupled_model}. This formulation of the model also allows us to utilize simulation methods such as the Gillespie algorithm to generate the trajectories of $\pmb{\eta}$ \cite{Gillespie1977}.

To compare the change in state space from continuous to discrete, we obtained an approximate SDE for the uncoupled model using a Kramers-Moyal expansion \cite{Gillespie1980, vankampen2007}. This relates the discrete transition rates $p_i$ and $q_i$ to the continuous parameters $v_i$ and $D_i$. The approximated SDE is a version of the general SDE in Eq. \eqref{eq:SDE_general} but where each $N_T$ therapies evolving independently of one another,
\begin{equation}\label{eq:SDE_uncoupled}
\begin{aligned}
    \mathrm{d}\eta_i =& \left(1-\mathcal{I}_{\chi_i}(t) \right)\left( -v_i \mathrm{d}t + D_i \mathrm{d} W_i(t) \right) \\
    &+ \mathcal{I}_{\chi_i}(t) (\eta_{i;0} - \eta_i),
\end{aligned}
\end{equation}
for $1 \leq i \leq N_T$. The details of the approximation and how $v_i$ and $D_i$ relates to the discrete transition rates $p_i$ and $q_i$ are found in Appendix \ref{AP:transrates}.

We recall that the RDT is computed by taking the time at which the overall therapy efficacy $\sqrt{\sum_{i=1}^{N_T} \eta_i^2}$ reaches below $\eta_{\text{min}} = 0.4$. For this model, this condition is imposed by removing states and transitions on the lattice that fall below this boundary, which consequently means that the transition matrix $\matr{W}$ is modified. The details for this modification are found in Appendix \ref{AP:transmatrix}.

We now obtain the mean RDT for a discrete space of states, following the method outlined in Ref.~\cite{Harunari2022, vanderMeer2022, sekimoto2021derivation}. Let $\eta$, $\eta_a$, and $\eta_i$ refer to states on the lattice. Specifically, $\eta_a$ is a state along the absorbing boundary, $\eta$ is the initial state $\eta \equiv (\eta_{1}^{(0)}, \ldots, \eta_{N_T}^{(0)})$, and $\eta_i$ is any state on the lattice that is neither $\eta$ nor $\eta_a$. The mean time at which the therapy efficacy reaches an absorbing state $\eta_a$ starting from the initial state $\eta$ at time $t$ is
\begin{equation}\label{eq:discmfpt}
        \langle T  \vert \eta \rangle = \sum_{\forall (\eta_i \to \eta_a)} \left[\matr{W}^\top\right]_{\eta_a,\eta_i} \left[(\matr{S}^{-1})^2 \right]_{\eta_i,\eta},
\end{equation}
where we sum over all single-step transitions that lead to the absorbing boundary $(\eta_i \to \eta_a), \forall \eta_i \neq \eta_a$. The matrix $\matr{S}$ is called the \textit{survival} matrix equal to $\matr{W}$ with transition elements $(\eta \to \eta_a)$ set to zero, as discussed in \cite{Harunari2022, vanderMeer2022, sekimoto2021derivation}.

\section{\label{sec:results1} Resistance development time as a function of therapy administration strategies}

First, we aim to determine the impact of multiple therapies on the overall dynamics, as it is known to induce non-linear effects in the overall efficacy \cite{Roemhild2022}. In this section, we investigate different parameter regions to examine how multiple therapies affect the mean RDT. 

\subsection{\label{sec:effect_of_parameters_result} Switching rates and number of therapies may lead to detrimental effects}
\begin{figure*}[!htb]
    \centering
    \includegraphics[width=0.9\textwidth]{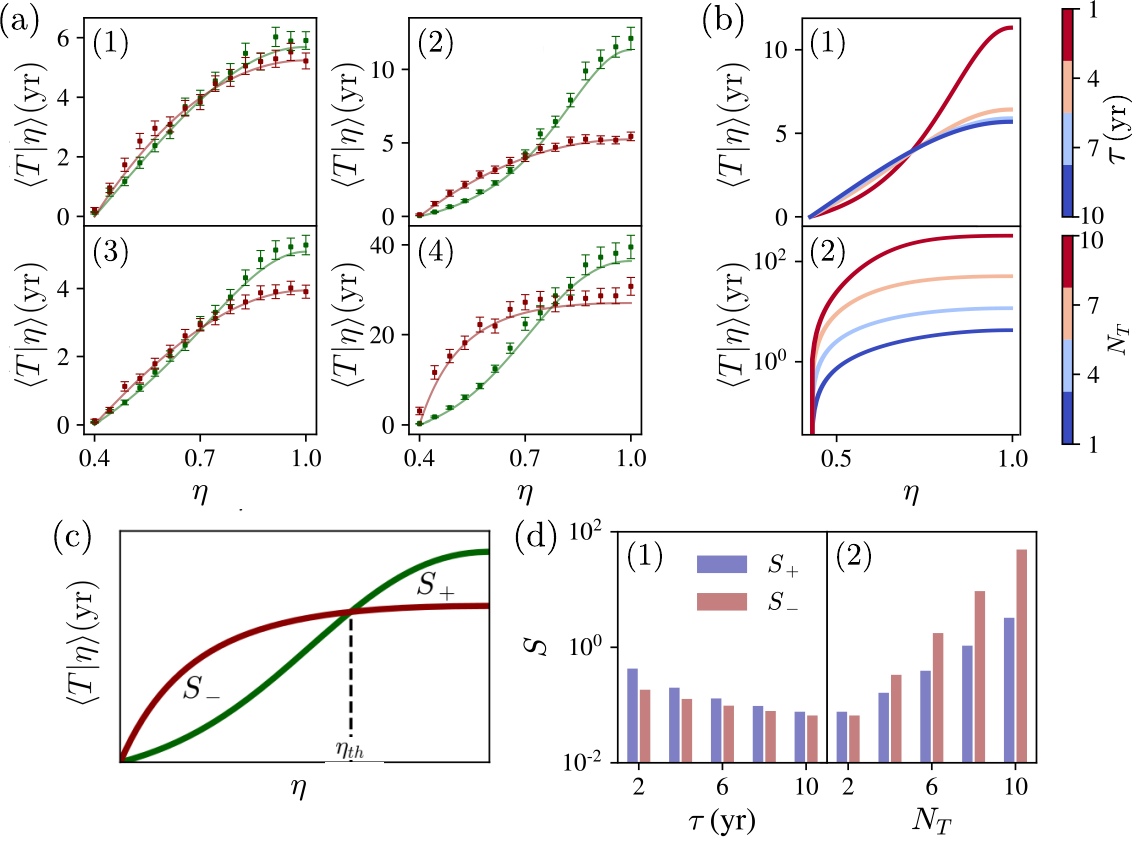}
    \caption{(a) Mean RDT with and without therapy resets (green and red respectively) as a function of initial therapy efficacy for (a.1) $\tau = 10$ yrs and $N_T = 2$, (a.2) $\tau = 1$ yr and $N_T = 2$, (a.3) $N_T = 1$ and $\tau = 3$ yrs, and (a.4) $N_T = 6$ and $\tau = 3$ yrs. Solid lines were obtained with Eq.~\eqref{eq:AFPT_coupled}, while dots and error bars are sample means and the standard error computed from $10^3$ Euler-Maruyama simulations of the process. (b) Mean RDT as a function of initial therapy efficacy for (b.1) different values of $\tau$ and (b.2) different values of $N_T$. (c) An illustration of the curves for the mean RDT as a function of initial therapy efficacy with and without therapy switching, highlighting the intersection of the curves $\eta_{th}$, which define the boundaries of $S_+$ and $S_-$. (d) Differences between the areas under the curve before (red) and after (blue) $\eta_{th}$ (d.1) as a function of $\tau$ with $N_T=2$ and (d.2) as a function of $N_T$ and $\tau= 3$ yrs. Parameters: $v = -8\cdot10^{-5}$ days$^{-1}$ and $D=10^{-4}$ days$^{-1}$.}
    \label{fig:effect_of_parameters}
\end{figure*}

In Fig. \ref{fig:effect_of_parameters}, we plot the mean RDT conditioned on the initial therapy efficacy, $\langle T \vert \eta \rangle$, using the coupled model for therapies with and without drug switching. Figures \hyperref[fig:effect_of_parameters]{3a.1} and \hyperref[fig:effect_of_parameters]{3a.2} show that there is an increase in the overall differences between the curves of switching and non-switching as $\tau$ decreases. Similarly, Figs. \hyperref[fig:effect_of_parameters]{3a.3} and \hyperref[fig:effect_of_parameters]{3a.4} show an increase in the difference between the two curves as $N_T$ increases. This increase is consistent for any $\tau$ or $N_T$, as shown by the analytical results in Figs. \hyperref[fig:effect_of_parameters]{3b.1} and \hyperref[fig:effect_of_parameters]{3b.2}.

We can always identify a threshold in the initial therapy efficacy, $\eta_{th}$, that marks the transition between a region where therapy switches are beneficial ($\eta>\eta_{th}$), in the sense that therapy switches tend to increase the RDT, and another region in which therapy switches are detrimental ($\eta<\eta_{th}$). To quantify the relevance of the beneficial and detrimental regions, the area between the mean RDT curves with and without therapy switches is computed for $\eta_{th} \leq \eta < 1$, which we call $S_+$, and $0 \leq \eta < \eta_{th}$, called $S_-$. The intersection $\eta_{th}$ together with the areas $S_-$ and $S_+$ are illustrated in Fig. \hyperref[fig:effect_of_parameters]{3c}. Figures \hyperref[fig:effect_of_parameters]{3d.1} and \hyperref[fig:effect_of_parameters]{3d.2} show that $S_+$ and $S_-$ decrease with increasing $\tau$, but the opposite happens for increasing $N_T$. We also note that  $S_-$ increases faster than $S_+$ when increasing $N_T$. This suggests that patients undergoing multi-drug therapies are at risk of experiencing significant detrimental effects, in the sense of a decrease in their mean RDT, due to therapy switches.

The remaining parameters of the coupled model, $v$ and $D$, do not introduce any new phenomenology beyond what is already present in $\tau$ and $N_T$. These parameters are discussed further in Appendix~\ref{sec:effect_of_parameters}.

\subsection{\label{sec:effect_of_decoupling} Determining the probability of detrimental therapy change protocols}

\begin{figure}[!htb]
    \centering
    \includegraphics[width = \linewidth]{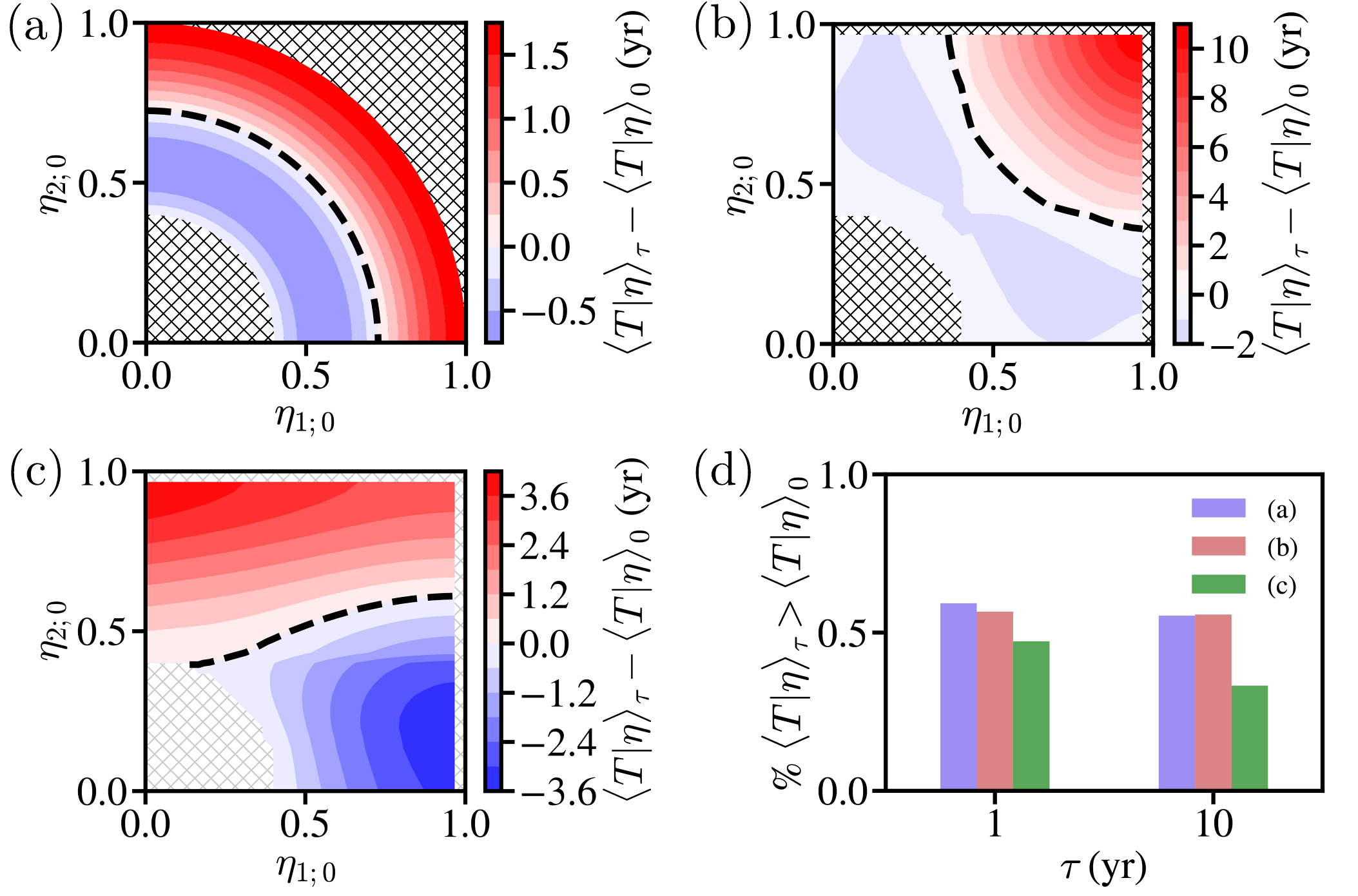}
        \caption{Difference between the mean RDT with therapy switching and without therapy switching as a function of initial therapy efficacy for (a) the coupled continuous model, (b) uncoupled discrete model with switching allowed for both therapies, and (c) uncoupled discrete model with switching allowed only for 1 drug $\eta_1$. Dashed lines indicate the region where the mean RDTs coincide. (d) The fraction occupied by the initial therapy states that the mean RDT with switching is higher than without switching as a function of $\tau$. Parameters: $\tau = 3$, $N_T = 2$, $v = -8\cdot10^{-5}$ days$^{-1}$ and $D=10^{-4}$ days$^{-1}$.}
\label{fig:effect_of_decoupling}
\end{figure}

Areas where mean RDT is higher for a therapy with or without switching are also observed for the uncoupled model. However, uncoupling implies that the therapies evolve independently of one another. The difference of the mean RDT with and without therapy switching is calculated for all the possible combinations of initial therapy efficacy for $N_T = 2$ in Fig. \ref{fig:effect_of_decoupling}. In this figure, the explicit beneficial and detrimental regions is illustrated in 2D space.

Figure \hyperref[fig:effect_of_decoupling]{4a} is consistent with the results in Fig. \ref{fig:effect_of_parameters}, where the region where therapy switching is more beneficial dominates when it is farther from the absorbing boundary. The same is observed in Fig. \hyperref[fig:effect_of_decoupling]{4b} but for the uncoupled model solved using Eq. \eqref{eq:discmfpt}. The uncoupled model allows for switching to be done for only one of the two therapies, as shown in Fig. \hyperref[fig:effect_of_decoupling]{4c}, and the beneficial or detrimental regions now yield a non-trivial gradient. This gradient allows us to obtain insights of where the beneficial and detrimental regions are by using the initial efficacy of the non-switching therapy. 

Figure \hyperref[fig:effect_of_decoupling]{4d} shows that the percentage of space occupied by the beneficial region. This quantity is of interest because it refers to the probability that the therapy design is beneficial for patients assuming the initial therapy efficacies $\pmb{\eta}$ are uniformly distributed. The beneficial region does not vary much as the switching rate is increased for both the coupled model and for the case of the uncoupled model where both therapies have equal switching rates. However, this is not the case for switching only one of the two therapies. The symmetry of the therapy efficacy space is relevant to estimate a priori the probability that a therapy design puts patients at risk.

\section{\label{sec:results2} Limiting therapies reveals optimal strategies}

With the current form of the model, a trivial solution to maximizing the RDT would be an infinite reset rate. However, this infinite switching is not feasible in a realistic clinical scenario because the number of therapies available to a patient is limited. For this reason, in this section, we study the effects of imposing limits or costs to a therapy switch. Additionally, we consider variations in the initial therapy efficacy for different therapies.

\subsection{\label{sec:limited_resetting_rate} Impact of stochasticity in the initial efficacy}

\begin{figure}[!htb]
    \centering
    \includegraphics[scale=0.6]{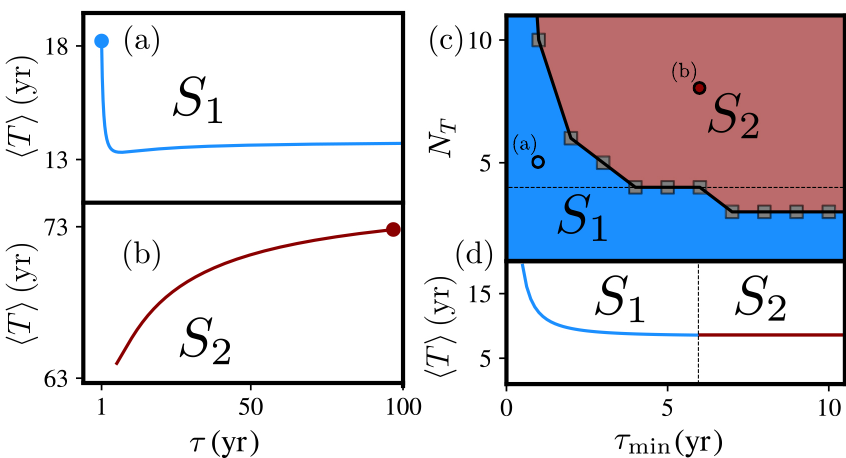}
    \caption{Unconditional mean RDT as a function of $\tau$ for (a) a representative trajectory where the maximum RDT is at $\tau_{min}$ ($S_1$) for $N_T = 5$, $\tau_{min} = 1$ and (b) a representative trajectory where the maximum RDT is not at $\tau_{min}$ ($S_2$) for $N_T = 8$, $\tau_{min} = 6$. Solid dots indicate the maximum RDT on each curve. (c) Phase diagram separating the regions for cases where the maximum RDT is and is not at $\tau_{min}$ as a function of $N_T$ and $\tau_{min}$. Solid dots on the phase diagram correspond to the representative curves in (a) and (b). (d) A transect of the phase diagram for $N_T=4$, corresponding to the horizontal dotted line in (c).}
    \label{fig:critical_dimension}
\end{figure}

Our first approach to model limited resources in therapies is to include a minimum frequency between therapy switches ($\tau_\text{min}$). We also consider that the efficacy right after therapy switches is a random variable with a complex behavior reflecting the interplay of drug-drug and patient-drugs. For simplicity, we assume that the therapy efficacy after switching is sampled from a uniform distribution on the entire therapy efficacy space. Thus, we define an unconditional mean RDT $\langle T \rangle$, which is the conditional mean RDT for the coupled model in Eq. \eqref{eq:AFPT_coupled} integrated over all possible values that the therapy efficacy will take after switching,
\begin{equation}\label{eq:total_average}
        \langle  T  \rangle =\frac{1}{\eta_{\max}^{N_T}-\eta_{\min}^{N_T}} \int d{\eta} \, \eta^{N_T-1}\, \langle T  \vert \eta \rangle .
\end{equation}
This expression is averaged over the total effective volume of the efficacy space $\eta_{\max}^{N_T}-\eta_{\min}^{N_T}$ for the coupled model, and the $\eta^{N_T-1}$ term in the integral accounts for the Jacobian of the change from Cartesian to spherical coordinates. Further details may be found in Appendix \ref{ap:uncondRDT}.

Within this setting, and for fixed number of simultaneous drugs $N_T$, there is an optimal value for the average frequency between drug switches $\tau$ that maximizes the mean RDT. Furthermore, this optimal value for $\tau$ changes abruptly depending on $N_T$ and $\tau_\text{min}$. Indeed, we can define a phase $S_1$, in which the optimal strategy is to apply as many therapy changes as possible, since $\tau=\tau_\text{min}$. In contrast, $S_2$ is another optimal strategy in which no therapy changes are applied, since $\tau\to\infty$.

Figures~\hyperref[fig:critical_dimension]{5a} and~\hyperref[fig:critical_dimension]{5b} show the mean RDT curves that correspond to the two phases $S_1$ and $S_2$. The curves transition from $S_1$ in Fig.~\hyperref[fig:critical_dimension]{5a} to $S_2$ in Fig.~\hyperref[fig:critical_dimension]{5b} by increasing the allowed minimum average therapy switching frequency from $\tau_{\min} = 1$ and $N_T = 5$ to $\tau_{\min} = 6$ and $N_T = 8$. In Fig.~\hyperref[fig:critical_dimension]{5c}, we show the boundary separating the phases as a function of $N_T$ and $\tau_{\min}$, with a transect of this phase diagram is shown in Fig.~\hyperref[fig:critical_dimension]{5d}. Therapy strategies in the phase $S_2$ may be more economical than those in $S_1$, since no drug switches are needed. Hence, for fixed $\tau_\text{min}$, increasing the number of simultaneous therapies $N_T$ to enter the phase $S_2$ could result in a more efficient strategy regarding both the maximization of the mean RDT and the minimization of the therapy costs. 

\subsection{\label{sec:limited_resetting} Fixing limits and costs to switching therapies}

\begin{figure*}[!htb]
    \centering
    \includegraphics[width=\textwidth]{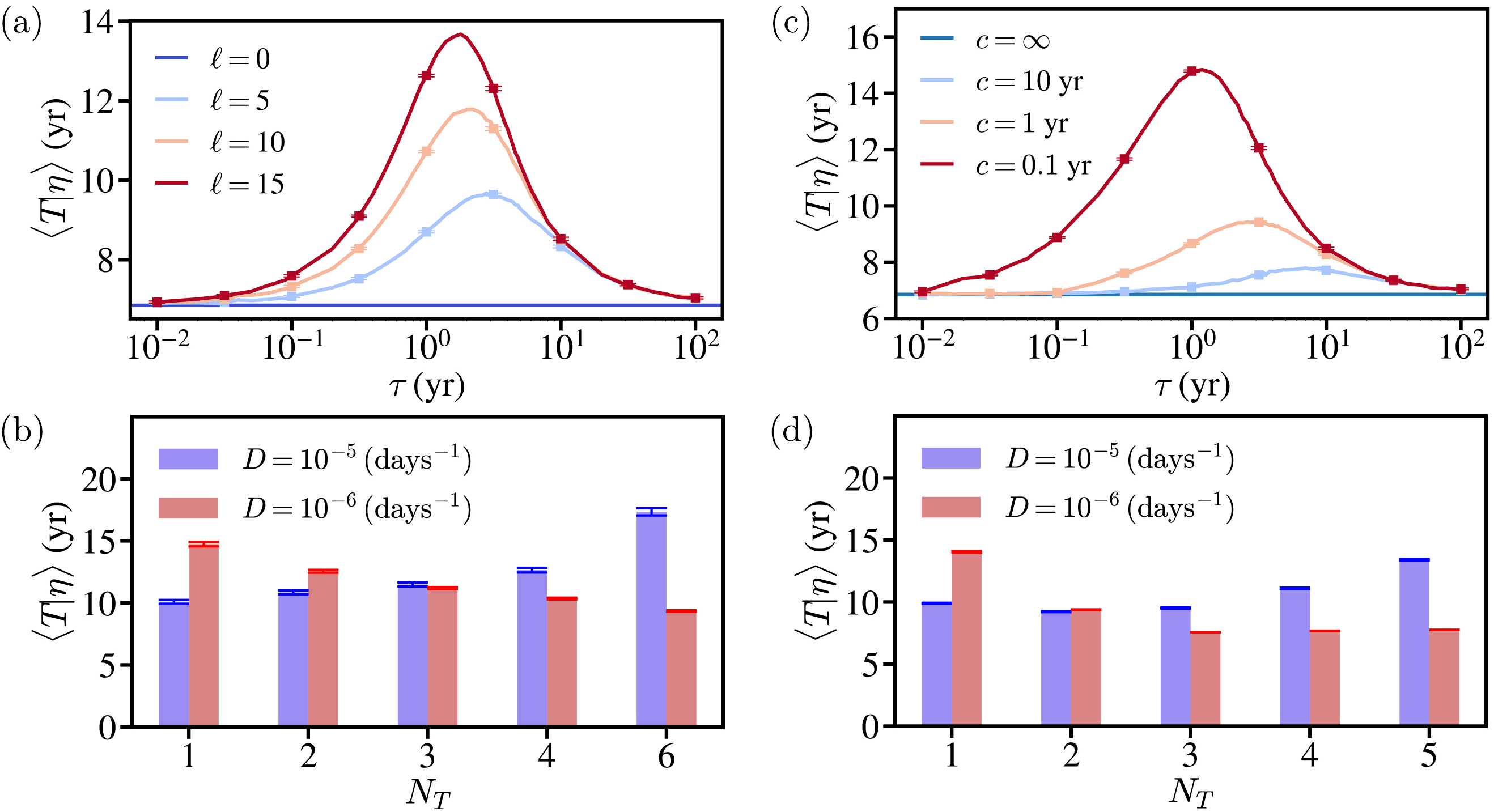}
    \caption{Mean RDT with limited switching (a) as a function of $\tau$ and $\ell$ using the uncoupled discrete model, and (b) as a function of $N_T$, two values of $D$, $A_T = 12$ and $\ell$ according to Eq. \eqref{eq:lconstraint} using the coupled continuous model. Mean RDT with costed switching (c) as a function of $\tau$ and $c$ using the uncoupled discrete model, and (d) as a function of $N_T$ and two values of $D$ using a cost function shown in Eq. \eqref{eq:costfcncoupled} using the coupled continuous model. Parameters: $v = -8 \cdot 10^{-5} \text{ days}^{-1}$ and $10^6$ simulations.}
    \label{fig:limited_resets}
\end{figure*}

Our next step is to simulate an explicit limit to the amount of therapy switches, let $\ell$ be the number of allowed switches, so that switching is no longer allowed after the $\ell$\textsuperscript{th} switch. We identify two methods of optimizing the mean RDT: first by varying the therapy change rate, and second by varying the number of simultaneous therapies.

We begin by fixing the number of simultaneous therapies to $N_T = 2$ and study the mean RDT as a function of the therapy switching rate for different number of allowed therapies $\ell$. The mean RDT is obtained using simulations of the uncoupled discrete model with the Gillespie algorithm using the transition matrix $\matr{W}$ as defined in Eq. \eqref{eq:gridmastereqn}. In Fig. \hyperref[fig:limited_resets]{6a}, we reveal a non-monotonic behavior for the mean RDT as the switching rate is varied. This suggests that given a limited number of therapies available, there is an optimal therapy switching rate that will maximize the mean RDT.

Next, we fix the switch rate $\tau$ while varying the number of simultaneous therapies $N_T$. We further constrain this by considering a total number of therapies $A_T$ and that the number of therapies consumed at the $\ell$\textsuperscript{th} switch is equal to the number of simultaneous therapies $N_T$. For example, given $A_T = 12$ total therapies available, for $N_T = 1$ simultaneous therapies then $\ell = 11$ switches are possible, for $N_T = 2$ we have $\ell = 5$, and so on. In general, for any number of $N_T$ and $A_T$, the number of allowed switches is
\begin{equation}\label{eq:lconstraint}
    \ell = \frac{A_T}{N_T}-1.
\end{equation}
The mean RDT for this version of limited switching is computed using simulations of the coupled continuous model using the Euler-Maruyama algorithm of the SDE in Eq. \eqref{eq:SDE_coupled}. Here we see a change in the dependence of the mean RDT with $D$: from an increasing behavior for a larger $D$ to a decreasing behavior for a smaller $D$. In the model, the diffusion constant $D$ controls how much therapy efficacy fluctuates over time and therefore it should grow with the mutation rate of the infecting pathogen. Figure \hyperref[fig:limited_resets]{6b} suggests that pathogens that mutate slower (i.e. smaller $D$) benefit from fewer simultaneous therapies and more limited switches, while the opposite is true for faster mutating pathogens.

Next, we consider that the therapy switching rate increases with the number of switches, reflecting a cost with each switch \cite{Sunil2023, debruyne2023}. Let $\gamma$ be the number of therapies that have been administered at a certain time. We replace the therapy switching rate with a function,
\begin{equation}\label{eq:costfcn}
    C(\gamma) = \frac{1}{\tau} \exp \left ( -\frac{c}{\tau} \gamma\right),
\end{equation}
where $c$ is a cost parameter that represents the average time taken for new therapies to be made available to the host, and $\tau$ is the time between therapy switches without cost. This function ensures that successive switches in the therapy $\gamma \to \infty$ diminishes the therapy switching rate $1/\tau \to 0$.

We observe similarities in the simulations of Fig. \hyperref[fig:limited_resets]{6a} to Fig. \hyperref[fig:limited_resets]{6c} for the uncoupled discrete model, and Fig. \hyperref[fig:limited_resets]{6b} to Fig. \hyperref[fig:limited_resets]{6d} for the coupled continuous model. Lower cost parameters $c$ allow for more switching in a shorter amount of time emulating a limit in the therapies. Fixing the number of simultaneous therapies to $N_T = 2$, this results to an existence of an optimal switching rate that maximizes the RDT, as shown in Fig. \hyperref[fig:limited_resets]{6c} taken from simulations using the Gillespie algorithm.

Furthermore, in Fig. \hyperref[fig:limited_resets]{6d}, we show  the RDT for two values of $D$ for the coupled model. The cost function is altered for the coupled model, since we must consider that each therapy switch changes all $N_T$ therapies at the same time. Lastly, we assume that increasing the number of simultaneous therapies in the coupled model becomes  more costly for each therapy switch, hence we assume that the cost parameter is dependent on $N_T$, using~$c = 10^{N_T-1}$. This yields a modified cost function for the coupled model given by
\begin{equation}\label{eq:costfcncoupled}
    C(\gamma) = \frac{1}{\tau} \exp \left ( -\frac{10^{N_T-1}}{\tau} \gamma\right).
\end{equation}
Figure \hyperref[fig:limited_resets]{6d} suggests that more simultaneous therapies, although costlier, are beneficial for more volatile infections with large $D$, and vice versa for small $D$.

\section{\label{sec:discussion} Discussion}

In this work, we have presented and characterized a stochastic model to study therapy administration strategies that aim to reduce drug resistance development. In particular, we have identified how therapy switches and the combination of therapies impact the resistance development time with and without restrictions on the drug availability. 

Our model has extended previous results on stochastic models of therapy administration~\cite{ramoso_stochastic_2020} by accounting for multiple therapies administered to the host which are known to induce non-linear changes in the overall therapy efficacy \cite{Roemhild2022}. We considered a therapy efficacy that scales the rate of infection in a host-pathogen model of HIV-1 dynamics allowing us to study chronic diseases \cite{Rong2009, Callaway2002}. Typically, therapy efficacy is included in the models as a constant rate or a functional form for the rate of infection and mutation \cite{Pinheiro2021, Rong2009,Callaway2002, Sharomi2008, Baral2019}. However, drug resistance can be modelled as a stochastic process \cite{prosperi2008}, allowing our model to account for temporal changes in therapy efficacy due to the noisy effects of pathogenic evolution. We have presented two models to study this phenomenon, namely the coupled continuous model and the uncoupled discrete model. The coupled continuous model eases the analytical study, but it has more limitations. On the other hand, the uncoupled discrete model allows for a more realistic study of the process, but it is computationally expensive.

In cases of therapies with higher initial therapy efficacies, both models have shown that increasing either the number of simultaneous therapies or the switching rate increases the mean RDT, consistent with the literature on combination and sequential therapies \cite{band2019antibiotic, Andersson2019, Nicoloff2019, Roemhild2019, Rosenkilde2019, Imamovic2013, Batra2021, oette2012}. The models identified conditions in terms of initial efficacy, for therapy switching to be beneficial or detrimental to patient outcomes. The detrimental effect is observed in case studies where adverse effects have resulted from drug switching \cite{Kirby2019, Sochman2005}. When the switching rates are equal for all therapies, both models show similar results. However, with the discrete model, it is possible to study asymmetries in the switching rates that yield qualitative differences in the beneficial and detrimental regions. This shows that the uncoupled discrete model is useful when each therapy has a different impact on the patient.

As an extension to the model, restrictions in terms of the availability of the therapy were included. This allows for the study of the optimal allocation of resources and treatment, which has been studied previously for chronic illnesses such as HIV \cite{AVANCENA20201509, Brandeau2008, Duwal2015}. The different restrictions that we have considered are a maximum therapy switching rate, limitations in the available therapies, and costs to therapy switching. These restrictions avoid unrealistic strategies such as an infinite therapy switching rate.

With the constraint of a maximum therapy switching rate imposed, we decided to study the uncertainty in therapy efficacy after a switch. We assumed the efficacy after a switch is a random variable and computed the mean RDT as a function of the switching rate. In this scenario, the optimal strategy is either switch therapies as fast as allowed or to not switch at all. The phase diagram for these two strategies, in terms of the number of therapies and the maximum switching rate, suggests that increasing the number of simultaneous therapies may lead to a scenario where switching therapies is no longer necessary, potentially reducing costs while maximizing the mean RDT.

Furthermore, the model has been constrained to consider only a limited number of therapies available or an increasing cost for each switch. In this setting, we showed that there is an optimal therapy switching rate that maximizes the RDT. It has been shown that if the costs for treatment and diagnostic check-ups are significant, then therapy switching without these diagnostic check-ups may fare better for the patient compared to evenly spaced diagnostic check-ups \cite{Duwal2015}. Our results for limited and costed therapy switching suggest the existence of optimal periods of when patients may undergo diagnostic check-ups with a physician to reduce the risk of therapy failure while minimizing costs.

In addition to the therapy switching rate, another controllable variable is the number of simultaneous therapies. In this case, we found that for high values of the diffusion constant, the mean RDT increases with the number of therapies. On the other hand, lower diffusion constants decrease the mean RDT as a function of the number of therapies. Note that the diffusion constant varies with each pathogen since it is correlated to its mutation rate. Therefore, our results suggest that for smaller rates of mutation, less simultaneous therapies are beneficial, while the opposite is true for larger mutation rates.

Currently our work only considers a model of drug interaction where drugs are considered to have independent modes of action within the host, hence the overall therapy efficacy in the infection rate is expressed as a product $\prod_{i=1}^{N_T}(1-\eta_i)$ \cite{Greco1995-ei, BLISS1939}. However, different modes of drug interactions may also suppress or amplify the overall therapy efficacy \cite{Andersson2019}, which may potentially change the form of the therapy efficacy. These interactions may also be simultaneous or sequential \cite{Roemhild2022}, potentially leading to changes in the optimal reset rates that have been identified. Furthermore, data obtained from empirical models of drug resistance may be used to inform the model \cite{chen2024drug, lamirande2024first}. While our work aims to study antimicrobial resistance, the model can easily be adapted to study drug resistance and drug switching in chronic or long-term illnesses without a pathogenic vector, such as diabetes \cite{Lebovitz2004}, COPD \cite{Braido2015}, and hypertension \cite{Wong2013, Vclavk2014, Athyros2010}.

Taken together, our model of therapy efficacy shows that the brute force strategy of increasing the rate of therapy switches and the number of drugs in the therapy might not result in better outcomes. In contrast, we can identify balanced optimal therapy choices depending on resource constraints that physicians designing therapies might face. While measuring the precise interactions between drugs, pathogens, and hosts can be complex, we have demonstrated that pathogen properties---such as the symmetries of phenotype landscapes and mutation rates, which can be assessed through empirical methods \cite{papkou2023rugged}---provide valuable insights for combating antimicrobial resistance.

\begin{acknowledgments}
This work was supported by the Ruth \& Arthur Scherbarth Foundation, the UniBE ID Grant 2021, the Agencia Estatal de Investigación (AEI, MCI, Spain) MCIN/AEI/10.13039/501100011033 and Fondo Europeo de Desarrollo Regional (FEDER, UE) under Project APASOS (PID2021$\textbf{-}$ 122256NB$\textbf{-}$C21$/$C22) and the María de Maeztu Program for units of Excellence in $\text{R\&D}$, grant CEX2021$\text{-}$ 001164$\text{-}$M.
\end{acknowledgments}

\bibliography{main}

\providecommand{\noopsort}[1]{}\providecommand{\singleletter}[1]{#1}%
\begin{thebibliography}{77}%
\makeatletter
\providecommand \@ifxundefined [1]{%
 \@ifx{#1\undefined}
}%
\providecommand \@ifnum [1]{%
 \ifnum #1\expandafter \@firstoftwo
 \else \expandafter \@secondoftwo
 \fi
}%
\providecommand \@ifx [1]{%
 \ifx #1\expandafter \@firstoftwo
 \else \expandafter \@secondoftwo
 \fi
}%
\providecommand \natexlab [1]{#1}%
\providecommand \enquote  [1]{``#1''}%
\providecommand \bibnamefont  [1]{#1}%
\providecommand \bibfnamefont [1]{#1}%
\providecommand \citenamefont [1]{#1}%
\providecommand \href@noop [0]{\@secondoftwo}%
\providecommand \href [0]{\begingroup \@sanitize@url \@href}%
\providecommand \@href[1]{\@@startlink{#1}\@@href}%
\providecommand \@@href[1]{\endgroup#1\@@endlink}%
\providecommand \@sanitize@url [0]{\catcode `\\12\catcode `\$12\catcode `\&12\catcode `\#12\catcode `\^12\catcode `\_12\catcode `\%12\relax}%
\providecommand \@@startlink[1]{}%
\providecommand \@@endlink[0]{}%
\providecommand \url  [0]{\begingroup\@sanitize@url \@url }%
\providecommand \@url [1]{\endgroup\@href {#1}{\urlprefix }}%
\providecommand \urlprefix  [0]{URL }%
\providecommand \Eprint [0]{\href }%
\providecommand \doibase [0]{https://doi.org/}%
\providecommand \selectlanguage [0]{\@gobble}%
\providecommand \bibinfo  [0]{\@secondoftwo}%
\providecommand \bibfield  [0]{\@secondoftwo}%
\providecommand \translation [1]{[#1]}%
\providecommand \BibitemOpen [0]{}%
\providecommand \bibitemStop [0]{}%
\providecommand \bibitemNoStop [0]{.\EOS\space}%
\providecommand \EOS [0]{\spacefactor3000\relax}%
\providecommand \BibitemShut  [1]{\csname bibitem#1\endcsname}%
\let\auto@bib@innerbib\@empty
\bibitem [{\citenamefont {Murray}\ and\ \citenamefont {et~al.}(2022)}]{Murray2022}%
  \BibitemOpen
  \bibfield  {author} {\bibinfo {author} {\bibfnamefont {C.~J.~L.}\ \bibnamefont {Murray}}\ and\ \bibinfo {author} {\bibnamefont {et~al.}},\ }\href {https://doi.org/10.1016/s0140-6736(21)02724-0} {\bibfield  {journal} {\bibinfo  {journal} {The Lancet}\ }\textbf {\bibinfo {volume} {399}},\ \bibinfo {pages} {629} (\bibinfo {year} {2022})}\BibitemShut {NoStop}%
\bibitem [{\citenamefont {Liebenberg}\ \emph {et~al.}(2022)\citenamefont {Liebenberg}, \citenamefont {Gordhan},\ and\ \citenamefont {Kana}}]{liebenberg2022drug}%
  \BibitemOpen
  \bibfield  {author} {\bibinfo {author} {\bibfnamefont {D.}~\bibnamefont {Liebenberg}}, \bibinfo {author} {\bibfnamefont {B.~G.}\ \bibnamefont {Gordhan}},\ and\ \bibinfo {author} {\bibfnamefont {B.~D.}\ \bibnamefont {Kana}},\ }\bibfield  {journal} {\bibinfo  {journal} {Frontiers in Cellular and Infection Microbiology}\ }\textbf {\bibinfo {volume} {12}},\ \href {https://doi.org/10.3389/fcimb.2022.943545} {10.3389/fcimb.2022.943545} (\bibinfo {year} {2022})\BibitemShut {NoStop}%
\bibitem [{\citenamefont {Phillips}\ \emph {et~al.}(2017)\citenamefont {Phillips}, \citenamefont {Stover}, \citenamefont {Cambiano}, \citenamefont {Nakagawa}, \citenamefont {Jordan}, \citenamefont {Pillay}, \citenamefont {Doherty}, \citenamefont {Revill},\ and\ \citenamefont {Bertagnolio}}]{phillips2017impact}%
  \BibitemOpen
  \bibfield  {author} {\bibinfo {author} {\bibfnamefont {A.~N.}\ \bibnamefont {Phillips}}, \bibinfo {author} {\bibfnamefont {J.}~\bibnamefont {Stover}}, \bibinfo {author} {\bibfnamefont {V.}~\bibnamefont {Cambiano}}, \bibinfo {author} {\bibfnamefont {F.}~\bibnamefont {Nakagawa}}, \bibinfo {author} {\bibfnamefont {M.~R.}\ \bibnamefont {Jordan}}, \bibinfo {author} {\bibfnamefont {D.}~\bibnamefont {Pillay}}, \bibinfo {author} {\bibfnamefont {M.}~\bibnamefont {Doherty}}, \bibinfo {author} {\bibfnamefont {P.}~\bibnamefont {Revill}},\ and\ \bibinfo {author} {\bibfnamefont {S.}~\bibnamefont {Bertagnolio}},\ }\href {https://doi.org/10.1093/infdis/jix089} {\bibfield  {journal} {\bibinfo  {journal} {The Journal of Infectious Diseases}\ }\textbf {\bibinfo {volume} {215}},\ \bibinfo {pages} {1362} (\bibinfo {year} {2017})}\BibitemShut {NoStop}%
\bibitem [{\citenamefont {Drechsler}\ and\ \citenamefont {Powderly}(2002)}]{Drechsler2002}%
  \BibitemOpen
  \bibfield  {author} {\bibinfo {author} {\bibfnamefont {H.}~\bibnamefont {Drechsler}}\ and\ \bibinfo {author} {\bibfnamefont {W.~G.}\ \bibnamefont {Powderly}},\ }\href {https://doi.org/10.1086/343050} {\bibfield  {journal} {\bibinfo  {journal} {Clinical Infectious Diseases}\ }\textbf {\bibinfo {volume} {35}},\ \bibinfo {pages} {1219–1230} (\bibinfo {year} {2002})}\BibitemShut {NoStop}%
\bibitem [{\citenamefont {Vasan}\ \emph {et~al.}(2019)\citenamefont {Vasan}, \citenamefont {Baselga},\ and\ \citenamefont {Hyman}}]{vasan2019view}%
  \BibitemOpen
  \bibfield  {author} {\bibinfo {author} {\bibfnamefont {N.}~\bibnamefont {Vasan}}, \bibinfo {author} {\bibfnamefont {J.}~\bibnamefont {Baselga}},\ and\ \bibinfo {author} {\bibfnamefont {D.~M.}\ \bibnamefont {Hyman}},\ }\href {https://doi.org/10.1038/s41586-019-1730-1} {\bibfield  {journal} {\bibinfo  {journal} {Nature}\ }\textbf {\bibinfo {volume} {575}},\ \bibinfo {pages} {299} (\bibinfo {year} {2019})}\BibitemShut {NoStop}%
\bibitem [{\citenamefont {Catalano}\ \emph {et~al.}(2022)\citenamefont {Catalano}, \citenamefont {Iacopetta}, \citenamefont {Ceramella}, \citenamefont {Scumaci}, \citenamefont {Giuzio}, \citenamefont {Saturnino}, \citenamefont {Aquaro}, \citenamefont {Rosano},\ and\ \citenamefont {Sinicropi}}]{catalano2022multidrug}%
  \BibitemOpen
  \bibfield  {author} {\bibinfo {author} {\bibfnamefont {A.}~\bibnamefont {Catalano}}, \bibinfo {author} {\bibfnamefont {D.}~\bibnamefont {Iacopetta}}, \bibinfo {author} {\bibfnamefont {J.}~\bibnamefont {Ceramella}}, \bibinfo {author} {\bibfnamefont {D.}~\bibnamefont {Scumaci}}, \bibinfo {author} {\bibfnamefont {F.}~\bibnamefont {Giuzio}}, \bibinfo {author} {\bibfnamefont {C.}~\bibnamefont {Saturnino}}, \bibinfo {author} {\bibfnamefont {S.}~\bibnamefont {Aquaro}}, \bibinfo {author} {\bibfnamefont {C.}~\bibnamefont {Rosano}},\ and\ \bibinfo {author} {\bibfnamefont {M.~S.}\ \bibnamefont {Sinicropi}},\ }\href@noop {} {\bibfield  {journal} {\bibinfo  {journal} {Molecules}\ }\textbf {\bibinfo {volume} {27}},\ \bibinfo {pages} {616} (\bibinfo {year} {2022})}\BibitemShut {NoStop}%
\bibitem [{\citenamefont {Maltas}\ \emph {et~al.}(2024)\citenamefont {Maltas}, \citenamefont {Tadele}, \citenamefont {Durmaz}, \citenamefont {McFarland}, \citenamefont {Hinczewski},\ and\ \citenamefont {Scott}}]{maltas2024}%
  \BibitemOpen
  \bibfield  {author} {\bibinfo {author} {\bibfnamefont {J.}~\bibnamefont {Maltas}}, \bibinfo {author} {\bibfnamefont {D.~S.}\ \bibnamefont {Tadele}}, \bibinfo {author} {\bibfnamefont {A.}~\bibnamefont {Durmaz}}, \bibinfo {author} {\bibfnamefont {C.~D.}\ \bibnamefont {McFarland}}, \bibinfo {author} {\bibfnamefont {M.}~\bibnamefont {Hinczewski}},\ and\ \bibinfo {author} {\bibfnamefont {J.~G.}\ \bibnamefont {Scott}},\ }\href {https://doi.org/10.1103/PRXLife.2.023010} {\bibfield  {journal} {\bibinfo  {journal} {PRX Life}\ }\textbf {\bibinfo {volume} {2}},\ \bibinfo {pages} {023010} (\bibinfo {year} {2024})}\BibitemShut {NoStop}%
\bibitem [{\citenamefont {Naylor}\ \emph {et~al.}(2018)\citenamefont {Naylor}, \citenamefont {Atun}, \citenamefont {Zhu}, \citenamefont {Kulasabanathan}, \citenamefont {Silva}, \citenamefont {Chatterjee}, \citenamefont {Knight},\ and\ \citenamefont {Robotham}}]{Naylor2018}%
  \BibitemOpen
  \bibfield  {author} {\bibinfo {author} {\bibfnamefont {N.~R.}\ \bibnamefont {Naylor}}, \bibinfo {author} {\bibfnamefont {R.}~\bibnamefont {Atun}}, \bibinfo {author} {\bibfnamefont {N.}~\bibnamefont {Zhu}}, \bibinfo {author} {\bibfnamefont {K.}~\bibnamefont {Kulasabanathan}}, \bibinfo {author} {\bibfnamefont {S.}~\bibnamefont {Silva}}, \bibinfo {author} {\bibfnamefont {A.}~\bibnamefont {Chatterjee}}, \bibinfo {author} {\bibfnamefont {G.~M.}\ \bibnamefont {Knight}},\ and\ \bibinfo {author} {\bibfnamefont {J.~V.}\ \bibnamefont {Robotham}},\ }\bibfield  {journal} {\bibinfo  {journal} {Antimicrobial Resistance \& Infection Control}\ }\textbf {\bibinfo {volume} {7}},\ \href {https://doi.org/10.1186/s13756-018-0336-y} {10.1186/s13756-018-0336-y} (\bibinfo {year} {2018})\BibitemShut {NoStop}%
\bibitem [{\citenamefont {O'Neill}(2016)}]{ONEILLAMR}%
  \BibitemOpen
  \bibfield  {author} {\bibinfo {author} {\bibfnamefont {J.}~\bibnamefont {O'Neill}},\ }\href {https://amr-review.org/Publications.html} {\emph {\bibinfo {title} {Tackling drug-resistant infections globally: final report and recommendations}}}\ (\bibinfo  {publisher} {Government of the United Kingdom},\ \bibinfo {year} {2016})\BibitemShut {NoStop}%
\bibitem [{\citenamefont {Langford}\ \emph {et~al.}(2023)\citenamefont {Langford}, \citenamefont {So}, \citenamefont {Simeonova}, \citenamefont {Leung}, \citenamefont {Lo}, \citenamefont {Kan}, \citenamefont {Raybardhan}, \citenamefont {Sapin}, \citenamefont {Mponponsuo}, \citenamefont {Farrell} \emph {et~al.}}]{langford2023antimicrobial}%
  \BibitemOpen
  \bibfield  {author} {\bibinfo {author} {\bibfnamefont {B.~J.}\ \bibnamefont {Langford}}, \bibinfo {author} {\bibfnamefont {M.}~\bibnamefont {So}}, \bibinfo {author} {\bibfnamefont {M.}~\bibnamefont {Simeonova}}, \bibinfo {author} {\bibfnamefont {V.}~\bibnamefont {Leung}}, \bibinfo {author} {\bibfnamefont {J.}~\bibnamefont {Lo}}, \bibinfo {author} {\bibfnamefont {T.}~\bibnamefont {Kan}}, \bibinfo {author} {\bibfnamefont {S.}~\bibnamefont {Raybardhan}}, \bibinfo {author} {\bibfnamefont {M.~E.}\ \bibnamefont {Sapin}}, \bibinfo {author} {\bibfnamefont {K.}~\bibnamefont {Mponponsuo}}, \bibinfo {author} {\bibfnamefont {A.}~\bibnamefont {Farrell}}, \emph {et~al.},\ }\href@noop {} {\bibfield  {journal} {\bibinfo  {journal} {The Lancet Microbe}\ }\textbf {\bibinfo {volume} {4}},\ \bibinfo {pages} {e179} (\bibinfo {year} {2023})}\BibitemShut {NoStop}%
\bibitem [{\citenamefont {Baym}\ \emph {et~al.}(2016)\citenamefont {Baym}, \citenamefont {Stone},\ and\ \citenamefont {Kishony}}]{baym2016multidrug}%
  \BibitemOpen
  \bibfield  {author} {\bibinfo {author} {\bibfnamefont {M.}~\bibnamefont {Baym}}, \bibinfo {author} {\bibfnamefont {L.~K.}\ \bibnamefont {Stone}},\ and\ \bibinfo {author} {\bibfnamefont {R.}~\bibnamefont {Kishony}},\ }\href@noop {} {\bibfield  {journal} {\bibinfo  {journal} {Science}\ }\textbf {\bibinfo {volume} {351}},\ \bibinfo {pages} {aad3292} (\bibinfo {year} {2016})}\BibitemShut {NoStop}%
\bibitem [{\citenamefont {Tyers}\ and\ \citenamefont {Wright}(2019)}]{tyers2019drug}%
  \BibitemOpen
  \bibfield  {author} {\bibinfo {author} {\bibfnamefont {M.}~\bibnamefont {Tyers}}\ and\ \bibinfo {author} {\bibfnamefont {G.~D.}\ \bibnamefont {Wright}},\ }\href@noop {} {\bibfield  {journal} {\bibinfo  {journal} {Nature Reviews Microbiology}\ }\textbf {\bibinfo {volume} {17}},\ \bibinfo {pages} {141} (\bibinfo {year} {2019})}\BibitemShut {NoStop}%
\bibitem [{\citenamefont {Band}\ \emph {et~al.}(2019)\citenamefont {Band}, \citenamefont {Hufnagel}, \citenamefont {Jaggavarapu}, \citenamefont {Sherman}, \citenamefont {Wozniak}, \citenamefont {Satola}, \citenamefont {Farley}, \citenamefont {Jacob}, \citenamefont {Burd},\ and\ \citenamefont {Weiss}}]{band2019antibiotic}%
  \BibitemOpen
  \bibfield  {author} {\bibinfo {author} {\bibfnamefont {V.~I.}\ \bibnamefont {Band}}, \bibinfo {author} {\bibfnamefont {D.~A.}\ \bibnamefont {Hufnagel}}, \bibinfo {author} {\bibfnamefont {S.}~\bibnamefont {Jaggavarapu}}, \bibinfo {author} {\bibfnamefont {E.~X.}\ \bibnamefont {Sherman}}, \bibinfo {author} {\bibfnamefont {J.~E.}\ \bibnamefont {Wozniak}}, \bibinfo {author} {\bibfnamefont {S.~W.}\ \bibnamefont {Satola}}, \bibinfo {author} {\bibfnamefont {M.~M.}\ \bibnamefont {Farley}}, \bibinfo {author} {\bibfnamefont {J.~T.}\ \bibnamefont {Jacob}}, \bibinfo {author} {\bibfnamefont {E.~M.}\ \bibnamefont {Burd}},\ and\ \bibinfo {author} {\bibfnamefont {D.~S.}\ \bibnamefont {Weiss}},\ }\href@noop {} {\bibfield  {journal} {\bibinfo  {journal} {Nature microbiology}\ }\textbf {\bibinfo {volume} {4}},\ \bibinfo {pages} {1627} (\bibinfo {year} {2019})}\BibitemShut {NoStop}%
\bibitem [{\citenamefont {Haas}\ \emph {et~al.}(2015)\citenamefont {Haas}, \citenamefont {Keiser}, \citenamefont {Balestre}, \citenamefont {Brown}, \citenamefont {Bissagnene}, \citenamefont {Chimbetete}, \citenamefont {Dabis}, \citenamefont {Davies}, \citenamefont {Hoffmann}, \citenamefont {Oyaro}, \citenamefont {Parkes-Ratanshi}, \citenamefont {Reynolds}, \citenamefont {Sikazwe}, \citenamefont {Wools-Kaloustian}, \citenamefont {Zannou}, \citenamefont {Wandeler},\ and\ \citenamefont {Egger}}]{Haas2015}%
  \BibitemOpen
  \bibfield  {author} {\bibinfo {author} {\bibfnamefont {A.~D.}\ \bibnamefont {Haas}}, \bibinfo {author} {\bibfnamefont {O.}~\bibnamefont {Keiser}}, \bibinfo {author} {\bibfnamefont {E.}~\bibnamefont {Balestre}}, \bibinfo {author} {\bibfnamefont {S.}~\bibnamefont {Brown}}, \bibinfo {author} {\bibfnamefont {E.}~\bibnamefont {Bissagnene}}, \bibinfo {author} {\bibfnamefont {C.}~\bibnamefont {Chimbetete}}, \bibinfo {author} {\bibfnamefont {F.}~\bibnamefont {Dabis}}, \bibinfo {author} {\bibfnamefont {M.-A.}\ \bibnamefont {Davies}}, \bibinfo {author} {\bibfnamefont {C.~J.}\ \bibnamefont {Hoffmann}}, \bibinfo {author} {\bibfnamefont {P.}~\bibnamefont {Oyaro}}, \bibinfo {author} {\bibfnamefont {R.}~\bibnamefont {Parkes-Ratanshi}}, \bibinfo {author} {\bibfnamefont {S.~J.}\ \bibnamefont {Reynolds}}, \bibinfo {author} {\bibfnamefont {I.}~\bibnamefont {Sikazwe}}, \bibinfo {author} {\bibfnamefont {K.}~\bibnamefont {Wools-Kaloustian}}, \bibinfo {author} {\bibfnamefont {D.~M.}\ \bibnamefont {Zannou}}, \bibinfo {author}
  {\bibfnamefont {G.}~\bibnamefont {Wandeler}},\ and\ \bibinfo {author} {\bibfnamefont {M.}~\bibnamefont {Egger}},\ }\href {https://doi.org/10.1016/s2352-3018(15)00087-9} {\bibfield  {journal} {\bibinfo  {journal} {The Lancet HIV}\ }\textbf {\bibinfo {volume} {2}},\ \bibinfo {pages} {e271–e278} (\bibinfo {year} {2015})}\BibitemShut {NoStop}%
\bibitem [{\citenamefont {Keiser}\ \emph {et~al.}(2009)\citenamefont {Keiser}, \citenamefont {MacPhail}, \citenamefont {Boulle}, \citenamefont {Wood}, \citenamefont {Schechter}, \citenamefont {Dabis}, \citenamefont {Sprinz},\ and\ \citenamefont {Egger}}]{Keiser2009}%
  \BibitemOpen
  \bibfield  {author} {\bibinfo {author} {\bibfnamefont {O.}~\bibnamefont {Keiser}}, \bibinfo {author} {\bibfnamefont {P.}~\bibnamefont {MacPhail}}, \bibinfo {author} {\bibfnamefont {A.}~\bibnamefont {Boulle}}, \bibinfo {author} {\bibfnamefont {R.}~\bibnamefont {Wood}}, \bibinfo {author} {\bibfnamefont {M.}~\bibnamefont {Schechter}}, \bibinfo {author} {\bibfnamefont {F.}~\bibnamefont {Dabis}}, \bibinfo {author} {\bibfnamefont {E.}~\bibnamefont {Sprinz}},\ and\ \bibinfo {author} {\bibfnamefont {M.}~\bibnamefont {Egger}},\ }\href {https://doi.org/10.1111/j.1365-3156.2009.02338.x} {\bibfield  {journal} {\bibinfo  {journal} {Tropical Medicine \& International Health}\ }\textbf {\bibinfo {volume} {14}},\ \bibinfo {pages} {1220–1225} (\bibinfo {year} {2009})}\BibitemShut {NoStop}%
\bibitem [{\citenamefont {Reynolds}\ \emph {et~al.}(2009)\citenamefont {Reynolds}, \citenamefont {Nakigozi}, \citenamefont {Newell}, \citenamefont {Ndyanabo}, \citenamefont {Galiwongo}, \citenamefont {Boaz}, \citenamefont {Quinn}, \citenamefont {Gray}, \citenamefont {Wawer},\ and\ \citenamefont {Serwadda}}]{Reynolds2009}%
  \BibitemOpen
  \bibfield  {author} {\bibinfo {author} {\bibfnamefont {S.~J.}\ \bibnamefont {Reynolds}}, \bibinfo {author} {\bibfnamefont {G.}~\bibnamefont {Nakigozi}}, \bibinfo {author} {\bibfnamefont {K.}~\bibnamefont {Newell}}, \bibinfo {author} {\bibfnamefont {A.}~\bibnamefont {Ndyanabo}}, \bibinfo {author} {\bibfnamefont {R.}~\bibnamefont {Galiwongo}}, \bibinfo {author} {\bibfnamefont {I.}~\bibnamefont {Boaz}}, \bibinfo {author} {\bibfnamefont {T.~C.}\ \bibnamefont {Quinn}}, \bibinfo {author} {\bibfnamefont {R.}~\bibnamefont {Gray}}, \bibinfo {author} {\bibfnamefont {M.}~\bibnamefont {Wawer}},\ and\ \bibinfo {author} {\bibfnamefont {D.}~\bibnamefont {Serwadda}},\ }\href {https://doi.org/10.1097/qad.0b013e3283262a78} {\bibfield  {journal} {\bibinfo  {journal} {AIDS}\ }\textbf {\bibinfo {volume} {23}},\ \bibinfo {pages} {697–700} (\bibinfo {year} {2009})}\BibitemShut {NoStop}%
\bibitem [{\citenamefont {Mayer}\ \emph {et~al.}(2002)\citenamefont {Mayer}, \citenamefont {Drechsler},\ and\ \citenamefont {Powderly}}]{mayer2002switching}%
  \BibitemOpen
  \bibfield  {author} {\bibinfo {author} {\bibfnamefont {K.~H.}\ \bibnamefont {Mayer}}, \bibinfo {author} {\bibfnamefont {H.}~\bibnamefont {Drechsler}},\ and\ \bibinfo {author} {\bibfnamefont {W.~G.}\ \bibnamefont {Powderly}},\ }\href@noop {} {\bibfield  {journal} {\bibinfo  {journal} {Clinical Infectious Diseases}\ }\textbf {\bibinfo {volume} {35}},\ \bibinfo {pages} {1219} (\bibinfo {year} {2002})}\BibitemShut {NoStop}%
\bibitem [{\citenamefont {Davidson}\ \emph {et~al.}(2010)\citenamefont {Davidson}, \citenamefont {Beardsell}, \citenamefont {Smith}, \citenamefont {Mandalia}, \citenamefont {Bower}, \citenamefont {Gazzard}, \citenamefont {Nelson},\ and\ \citenamefont {Stebbing}}]{DAVIDSON2010227}%
  \BibitemOpen
  \bibfield  {author} {\bibinfo {author} {\bibfnamefont {I.}~\bibnamefont {Davidson}}, \bibinfo {author} {\bibfnamefont {H.}~\bibnamefont {Beardsell}}, \bibinfo {author} {\bibfnamefont {B.}~\bibnamefont {Smith}}, \bibinfo {author} {\bibfnamefont {S.}~\bibnamefont {Mandalia}}, \bibinfo {author} {\bibfnamefont {M.}~\bibnamefont {Bower}}, \bibinfo {author} {\bibfnamefont {B.}~\bibnamefont {Gazzard}}, \bibinfo {author} {\bibfnamefont {M.}~\bibnamefont {Nelson}},\ and\ \bibinfo {author} {\bibfnamefont {J.}~\bibnamefont {Stebbing}},\ }\href {https://doi.org/https://doi.org/10.1016/j.antiviral.2010.03.001} {\bibfield  {journal} {\bibinfo  {journal} {Antiviral Research}\ }\textbf {\bibinfo {volume} {86}},\ \bibinfo {pages} {227} (\bibinfo {year} {2010})}\BibitemShut {NoStop}%
\bibitem [{\citenamefont {Kirby}\ \emph {et~al.}(2019)\citenamefont {Kirby}, \citenamefont {Allchorne}, \citenamefont {Appanna}, \citenamefont {Davey}, \citenamefont {Gledhill}, \citenamefont {Green}, \citenamefont {Greene},\ and\ \citenamefont {Rosario}}]{Kirby2019}%
  \BibitemOpen
  \bibfield  {author} {\bibinfo {author} {\bibfnamefont {M.~G.}\ \bibnamefont {Kirby}}, \bibinfo {author} {\bibfnamefont {P.}~\bibnamefont {Allchorne}}, \bibinfo {author} {\bibfnamefont {T.}~\bibnamefont {Appanna}}, \bibinfo {author} {\bibfnamefont {P.}~\bibnamefont {Davey}}, \bibinfo {author} {\bibfnamefont {R.}~\bibnamefont {Gledhill}}, \bibinfo {author} {\bibfnamefont {J.~S.~A.}\ \bibnamefont {Green}}, \bibinfo {author} {\bibfnamefont {D.}~\bibnamefont {Greene}},\ and\ \bibinfo {author} {\bibfnamefont {D.~J.}\ \bibnamefont {Rosario}},\ }\bibfield  {journal} {\bibinfo  {journal} {International Journal of Clinical Practice}\ }\textbf {\bibinfo {volume} {74}},\ \href {https://doi.org/10.1111/ijcp.13429} {10.1111/ijcp.13429} (\bibinfo {year} {2019})\BibitemShut {NoStop}%
\bibitem [{\citenamefont {Pinheiro}\ \emph {et~al.}(2021)\citenamefont {Pinheiro}, \citenamefont {Warsi}, \citenamefont {Andersson},\ and\ \citenamefont {L\"{a}ssig}}]{Pinheiro2021}%
  \BibitemOpen
  \bibfield  {author} {\bibinfo {author} {\bibfnamefont {F.}~\bibnamefont {Pinheiro}}, \bibinfo {author} {\bibfnamefont {O.}~\bibnamefont {Warsi}}, \bibinfo {author} {\bibfnamefont {D.~I.}\ \bibnamefont {Andersson}},\ and\ \bibinfo {author} {\bibfnamefont {M.}~\bibnamefont {L\"{a}ssig}},\ }\href {https://doi.org/10.1038/s41559-021-01397-0} {\bibfield  {journal} {\bibinfo  {journal} {Nature Ecology \& Evolution}\ }\textbf {\bibinfo {volume} {5}},\ \bibinfo {pages} {677–687} (\bibinfo {year} {2021})}\BibitemShut {NoStop}%
\bibitem [{\citenamefont {Andersson}\ \emph {et~al.}(2019)\citenamefont {Andersson}, \citenamefont {Nicoloff},\ and\ \citenamefont {Hjort}}]{Andersson2019}%
  \BibitemOpen
  \bibfield  {author} {\bibinfo {author} {\bibfnamefont {D.~I.}\ \bibnamefont {Andersson}}, \bibinfo {author} {\bibfnamefont {H.}~\bibnamefont {Nicoloff}},\ and\ \bibinfo {author} {\bibfnamefont {K.}~\bibnamefont {Hjort}},\ }\href {https://doi.org/10.1038/s41579-019-0218-1} {\bibfield  {journal} {\bibinfo  {journal} {Nature Reviews Microbiology}\ }\textbf {\bibinfo {volume} {17}},\ \bibinfo {pages} {479–496} (\bibinfo {year} {2019})}\BibitemShut {NoStop}%
\bibitem [{\citenamefont {Lee}\ \emph {et~al.}(2022)\citenamefont {Lee}, \citenamefont {Bonhoeffer},\ and\ \citenamefont {Penny}}]{Lee2022}%
  \BibitemOpen
  \bibfield  {author} {\bibinfo {author} {\bibfnamefont {T.}~\bibnamefont {Lee}}, \bibinfo {author} {\bibfnamefont {S.}~\bibnamefont {Bonhoeffer}},\ and\ \bibinfo {author} {\bibfnamefont {M.}~\bibnamefont {Penny}},\ }\href {https://doi.org/10.1016/j.rinp.2022.105181} {\bibfield  {journal} {\bibinfo  {journal} {Results in Physics}\ }\textbf {\bibinfo {volume} {34}},\ \bibinfo {pages} {105181} (\bibinfo {year} {2022})}\BibitemShut {NoStop}%
\bibitem [{\citenamefont {Rong}\ and\ \citenamefont {Perelson}(2009)}]{Rong2009}%
  \BibitemOpen
  \bibfield  {author} {\bibinfo {author} {\bibfnamefont {L.}~\bibnamefont {Rong}}\ and\ \bibinfo {author} {\bibfnamefont {A.~S.}\ \bibnamefont {Perelson}},\ }\href {https://doi.org/10.1371/journal.pcbi.1000533} {\bibfield  {journal} {\bibinfo  {journal} {{PLoS} Computational Biology}\ }\textbf {\bibinfo {volume} {5}},\ \bibinfo {pages} {e1000533} (\bibinfo {year} {2009})}\BibitemShut {NoStop}%
\bibitem [{\citenamefont {Duncan~Callaway}(2002)}]{Callaway2002}%
  \BibitemOpen
  \bibfield  {author} {\bibinfo {author} {\bibfnamefont {A.~P.}\ \bibnamefont {Duncan~Callaway}},\ }\href {https://doi.org/10.1006/bulm.2001.0266} {\bibfield  {journal} {\bibinfo  {journal} {Bulletin of Mathematical Biology}\ }\textbf {\bibinfo {volume} {64}},\ \bibinfo {pages} {29} (\bibinfo {year} {2002})}\BibitemShut {NoStop}%
\bibitem [{\citenamefont {Sharomi}\ and\ \citenamefont {Gumel}(2008)}]{Sharomi2008}%
  \BibitemOpen
  \bibfield  {author} {\bibinfo {author} {\bibfnamefont {O.}~\bibnamefont {Sharomi}}\ and\ \bibinfo {author} {\bibfnamefont {A.}~\bibnamefont {Gumel}},\ }\href {https://doi.org/10.1080/17513750701775599} {\bibfield  {journal} {\bibinfo  {journal} {Journal of Biological Dynamics}\ }\textbf {\bibinfo {volume} {2}},\ \bibinfo {pages} {323–345} (\bibinfo {year} {2008})}\BibitemShut {NoStop}%
\bibitem [{\citenamefont {Niewiadomska}\ \emph {et~al.}(2019)\citenamefont {Niewiadomska}, \citenamefont {Jayabalasingham}, \citenamefont {Seidman}, \citenamefont {Willem}, \citenamefont {Grenfell}, \citenamefont {Spiro},\ and\ \citenamefont {Viboud}}]{Niewiadomska2019}%
  \BibitemOpen
  \bibfield  {author} {\bibinfo {author} {\bibfnamefont {A.~M.}\ \bibnamefont {Niewiadomska}}, \bibinfo {author} {\bibfnamefont {B.}~\bibnamefont {Jayabalasingham}}, \bibinfo {author} {\bibfnamefont {J.~C.}\ \bibnamefont {Seidman}}, \bibinfo {author} {\bibfnamefont {L.}~\bibnamefont {Willem}}, \bibinfo {author} {\bibfnamefont {B.}~\bibnamefont {Grenfell}}, \bibinfo {author} {\bibfnamefont {D.}~\bibnamefont {Spiro}},\ and\ \bibinfo {author} {\bibfnamefont {C.}~\bibnamefont {Viboud}},\ }\bibfield  {journal} {\bibinfo  {journal} {BMC Medicine}\ }\textbf {\bibinfo {volume} {17}},\ \href {https://doi.org/10.1186/s12916-019-1314-9} {10.1186/s12916-019-1314-9} (\bibinfo {year} {2019})\BibitemShut {NoStop}%
\bibitem [{\citenamefont {Heesterbeek}\ \emph {et~al.}(2015)\citenamefont {Heesterbeek}, \citenamefont {Anderson}, \citenamefont {Andreasen}, \citenamefont {Bansal}, \citenamefont {De~Angelis}, \citenamefont {Dye}, \citenamefont {Eames}, \citenamefont {Edmunds}, \citenamefont {Frost}, \citenamefont {Funk}, \citenamefont {Hollingsworth}, \citenamefont {House}, \citenamefont {Isham}, \citenamefont {Klepac}, \citenamefont {Lessler}, \citenamefont {Lloyd-Smith}, \citenamefont {Metcalf}, \citenamefont {Mollison}, \citenamefont {Pellis}, \citenamefont {Pulliam}, \citenamefont {Roberts},\ and\ \citenamefont {Viboud}}]{Heesterbeek2015}%
  \BibitemOpen
  \bibfield  {author} {\bibinfo {author} {\bibfnamefont {H.}~\bibnamefont {Heesterbeek}}, \bibinfo {author} {\bibfnamefont {R.~M.}\ \bibnamefont {Anderson}}, \bibinfo {author} {\bibfnamefont {V.}~\bibnamefont {Andreasen}}, \bibinfo {author} {\bibfnamefont {S.}~\bibnamefont {Bansal}}, \bibinfo {author} {\bibfnamefont {D.}~\bibnamefont {De~Angelis}}, \bibinfo {author} {\bibfnamefont {C.}~\bibnamefont {Dye}}, \bibinfo {author} {\bibfnamefont {K.~T.~D.}\ \bibnamefont {Eames}}, \bibinfo {author} {\bibfnamefont {W.~J.}\ \bibnamefont {Edmunds}}, \bibinfo {author} {\bibfnamefont {S.~D.~W.}\ \bibnamefont {Frost}}, \bibinfo {author} {\bibfnamefont {S.}~\bibnamefont {Funk}}, \bibinfo {author} {\bibfnamefont {T.~D.}\ \bibnamefont {Hollingsworth}}, \bibinfo {author} {\bibfnamefont {T.}~\bibnamefont {House}}, \bibinfo {author} {\bibfnamefont {V.}~\bibnamefont {Isham}}, \bibinfo {author} {\bibfnamefont {P.}~\bibnamefont {Klepac}}, \bibinfo {author} {\bibfnamefont {J.}~\bibnamefont {Lessler}}, \bibinfo {author}
  {\bibfnamefont {J.~O.}\ \bibnamefont {Lloyd-Smith}}, \bibinfo {author} {\bibfnamefont {C.~J.~E.}\ \bibnamefont {Metcalf}}, \bibinfo {author} {\bibfnamefont {D.}~\bibnamefont {Mollison}}, \bibinfo {author} {\bibfnamefont {L.}~\bibnamefont {Pellis}}, \bibinfo {author} {\bibfnamefont {J.~R.~C.}\ \bibnamefont {Pulliam}}, \bibinfo {author} {\bibfnamefont {M.~G.}\ \bibnamefont {Roberts}},\ and\ \bibinfo {author} {\bibfnamefont {C.}~\bibnamefont {Viboud}},\ }\bibfield  {journal} {\bibinfo  {journal} {Science}\ }\textbf {\bibinfo {volume} {347}},\ \href {https://doi.org/10.1126/science.aaa4339} {10.1126/science.aaa4339} (\bibinfo {year} {2015})\BibitemShut {NoStop}%
\bibitem [{\citenamefont {Hillock}\ \emph {et~al.}(2022)\citenamefont {Hillock}, \citenamefont {Merlin}, \citenamefont {Turnidge},\ and\ \citenamefont {Karnon}}]{Hillock2022}%
  \BibitemOpen
  \bibfield  {author} {\bibinfo {author} {\bibfnamefont {N.~T.}\ \bibnamefont {Hillock}}, \bibinfo {author} {\bibfnamefont {T.~L.}\ \bibnamefont {Merlin}}, \bibinfo {author} {\bibfnamefont {J.}~\bibnamefont {Turnidge}},\ and\ \bibinfo {author} {\bibfnamefont {J.}~\bibnamefont {Karnon}},\ }\href {https://doi.org/10.1007/s40258-022-00728-x} {\bibfield  {journal} {\bibinfo  {journal} {Applied Health Economics and Health Policy}\ }\textbf {\bibinfo {volume} {20}},\ \bibinfo {pages} {479–486} (\bibinfo {year} {2022})}\BibitemShut {NoStop}%
\bibitem [{\citenamefont {Manrubia}(2012)}]{Manrubia2012}%
  \BibitemOpen
  \bibfield  {author} {\bibinfo {author} {\bibfnamefont {S.~C.}\ \bibnamefont {Manrubia}},\ }\href {https://doi.org/10.1016/j.coviro.2012.06.006} {\bibfield  {journal} {\bibinfo  {journal} {Current Opinion in Virology}\ }\textbf {\bibinfo {volume} {2}},\ \bibinfo {pages} {531–537} (\bibinfo {year} {2012})}\BibitemShut {NoStop}%
\bibitem [{\citenamefont {Papkou}\ \emph {et~al.}(2023)\citenamefont {Papkou}, \citenamefont {Garcia-Pastor}, \citenamefont {Escudero},\ and\ \citenamefont {Wagner}}]{papkou2023rugged}%
  \BibitemOpen
  \bibfield  {author} {\bibinfo {author} {\bibfnamefont {A.}~\bibnamefont {Papkou}}, \bibinfo {author} {\bibfnamefont {L.}~\bibnamefont {Garcia-Pastor}}, \bibinfo {author} {\bibfnamefont {J.~A.}\ \bibnamefont {Escudero}},\ and\ \bibinfo {author} {\bibfnamefont {A.}~\bibnamefont {Wagner}},\ }\href {https://doi.org/10.1126/science.adh3860} {\bibfield  {journal} {\bibinfo  {journal} {Science}\ }\textbf {\bibinfo {volume} {382}},\ \bibinfo {pages} {eadh3860} (\bibinfo {year} {2023})},\ \Eprint {https://arxiv.org/abs/https://www.science.org/doi/pdf/10.1126/science.adh3860} {https://www.science.org/doi/pdf/10.1126/science.adh3860} \BibitemShut {NoStop}%
\bibitem [{\citenamefont {Evans}\ and\ \citenamefont {Majumdar}(2011{\natexlab{a}})}]{evans_diffusion_2011}%
  \BibitemOpen
  \bibfield  {author} {\bibinfo {author} {\bibfnamefont {M.~R.}\ \bibnamefont {Evans}}\ and\ \bibinfo {author} {\bibfnamefont {S.~N.}\ \bibnamefont {Majumdar}},\ }\href {https://doi.org/10.1103/PhysRevLett.106.160601} {\bibfield  {journal} {\bibinfo  {journal} {Physical Review Letters}\ }\textbf {\bibinfo {volume} {106}},\ \bibinfo {pages} {160601} (\bibinfo {year} {2011}{\natexlab{a}})}\BibitemShut {NoStop}%
\bibitem [{\citenamefont {Evans}\ \emph {et~al.}(2020)\citenamefont {Evans}, \citenamefont {Majumdar},\ and\ \citenamefont {Schehr}}]{evans_stochastic_2020}%
  \BibitemOpen
  \bibfield  {author} {\bibinfo {author} {\bibfnamefont {M.~R.}\ \bibnamefont {Evans}}, \bibinfo {author} {\bibfnamefont {S.~N.}\ \bibnamefont {Majumdar}},\ and\ \bibinfo {author} {\bibfnamefont {G.}~\bibnamefont {Schehr}},\ }\href {https://doi.org/10.1088/1751-8121/ab7cfe} {\bibfield  {journal} {\bibinfo  {journal} {Journal of Physics A: Mathematical and Theoretical}\ }\textbf {\bibinfo {volume} {53}},\ \bibinfo {pages} {193001} (\bibinfo {year} {2020})}\BibitemShut {NoStop}%
\bibitem [{\citenamefont {Rotbart}\ \emph {et~al.}(2015)\citenamefont {Rotbart}, \citenamefont {Reuveni},\ and\ \citenamefont {Urbakh}}]{rotbart2015michaelis}%
  \BibitemOpen
  \bibfield  {author} {\bibinfo {author} {\bibfnamefont {T.}~\bibnamefont {Rotbart}}, \bibinfo {author} {\bibfnamefont {S.}~\bibnamefont {Reuveni}},\ and\ \bibinfo {author} {\bibfnamefont {M.}~\bibnamefont {Urbakh}},\ }\href@noop {} {\bibfield  {journal} {\bibinfo  {journal} {Physical Review E}\ }\textbf {\bibinfo {volume} {92}},\ \bibinfo {pages} {060101} (\bibinfo {year} {2015})}\BibitemShut {NoStop}%
\bibitem [{\citenamefont {Rold{\'a}n}\ \emph {et~al.}(2016)\citenamefont {Rold{\'a}n}, \citenamefont {Lisica}, \citenamefont {S{\'a}nchez-Taltavull},\ and\ \citenamefont {Grill}}]{roldan2016stochastic}%
  \BibitemOpen
  \bibfield  {author} {\bibinfo {author} {\bibfnamefont {{\'E}.}~\bibnamefont {Rold{\'a}n}}, \bibinfo {author} {\bibfnamefont {A.}~\bibnamefont {Lisica}}, \bibinfo {author} {\bibfnamefont {D.}~\bibnamefont {S{\'a}nchez-Taltavull}},\ and\ \bibinfo {author} {\bibfnamefont {S.~W.}\ \bibnamefont {Grill}},\ }\href@noop {} {\bibfield  {journal} {\bibinfo  {journal} {Physical Review E}\ }\textbf {\bibinfo {volume} {93}},\ \bibinfo {pages} {062411} (\bibinfo {year} {2016})}\BibitemShut {NoStop}%
\bibitem [{\citenamefont {Lisica}\ \emph {et~al.}(2016)\citenamefont {Lisica}, \citenamefont {Engel}, \citenamefont {Jahnel}, \citenamefont {Rold{\'a}n}, \citenamefont {Galburt}, \citenamefont {Cramer},\ and\ \citenamefont {Grill}}]{lisica2016mechanisms}%
  \BibitemOpen
  \bibfield  {author} {\bibinfo {author} {\bibfnamefont {A.}~\bibnamefont {Lisica}}, \bibinfo {author} {\bibfnamefont {C.}~\bibnamefont {Engel}}, \bibinfo {author} {\bibfnamefont {M.}~\bibnamefont {Jahnel}}, \bibinfo {author} {\bibfnamefont {{\'E}.}~\bibnamefont {Rold{\'a}n}}, \bibinfo {author} {\bibfnamefont {E.~A.}\ \bibnamefont {Galburt}}, \bibinfo {author} {\bibfnamefont {P.}~\bibnamefont {Cramer}},\ and\ \bibinfo {author} {\bibfnamefont {S.~W.}\ \bibnamefont {Grill}},\ }\href@noop {} {\bibfield  {journal} {\bibinfo  {journal} {Proceedings of the National Academy of Sciences}\ }\textbf {\bibinfo {volume} {113}},\ \bibinfo {pages} {2946} (\bibinfo {year} {2016})}\BibitemShut {NoStop}%
\bibitem [{\citenamefont {Bressloff}(2020)}]{bressloff2020modeling}%
  \BibitemOpen
  \bibfield  {author} {\bibinfo {author} {\bibfnamefont {P.~C.}\ \bibnamefont {Bressloff}},\ }\href@noop {} {\bibfield  {journal} {\bibinfo  {journal} {Journal of Physics A: Mathematical and Theoretical}\ }\textbf {\bibinfo {volume} {53}},\ \bibinfo {pages} {355001} (\bibinfo {year} {2020})}\BibitemShut {NoStop}%
\bibitem [{\citenamefont {Wang}\ \emph {et~al.}(2011)\citenamefont {Wang}, \citenamefont {Zhang}, \citenamefont {Xu},\ and\ \citenamefont {Wang}}]{wang2011quantifying}%
  \BibitemOpen
  \bibfield  {author} {\bibinfo {author} {\bibfnamefont {J.}~\bibnamefont {Wang}}, \bibinfo {author} {\bibfnamefont {K.}~\bibnamefont {Zhang}}, \bibinfo {author} {\bibfnamefont {L.}~\bibnamefont {Xu}},\ and\ \bibinfo {author} {\bibfnamefont {E.}~\bibnamefont {Wang}},\ }\href@noop {} {\bibfield  {journal} {\bibinfo  {journal} {Proceedings of the National Academy of Sciences}\ }\textbf {\bibinfo {volume} {108}},\ \bibinfo {pages} {8257} (\bibinfo {year} {2011})}\BibitemShut {NoStop}%
\bibitem [{\citenamefont {Redner}(2001)}]{redner2001guide}%
  \BibitemOpen
  \bibfield  {author} {\bibinfo {author} {\bibfnamefont {S.}~\bibnamefont {Redner}},\ }\href@noop {} {\emph {\bibinfo {title} {A guide to first-passage processes}}}\ (\bibinfo  {publisher} {Cambridge University Press},\ \bibinfo {year} {2001})\BibitemShut {NoStop}%
\bibitem [{\citenamefont {Evans}\ and\ \citenamefont {Majumdar}(2011{\natexlab{b}})}]{Evans2011}%
  \BibitemOpen
  \bibfield  {author} {\bibinfo {author} {\bibfnamefont {M.~R.}\ \bibnamefont {Evans}}\ and\ \bibinfo {author} {\bibfnamefont {S.~N.}\ \bibnamefont {Majumdar}},\ }\href {https://doi.org/10.1088/1751-8113/44/43/435001} {\bibfield  {journal} {\bibinfo  {journal} {Journal of Physics A: Mathematical and Theoretical}\ }\textbf {\bibinfo {volume} {44}},\ \bibinfo {pages} {435001} (\bibinfo {year} {2011}{\natexlab{b}})}\BibitemShut {NoStop}%
\bibitem [{\citenamefont {Pal}\ and\ \citenamefont {Prasad}(2019)}]{Pal2019}%
  \BibitemOpen
  \bibfield  {author} {\bibinfo {author} {\bibfnamefont {A.}~\bibnamefont {Pal}}\ and\ \bibinfo {author} {\bibfnamefont {V.~V.}\ \bibnamefont {Prasad}},\ }\bibfield  {journal} {\bibinfo  {journal} {Physical Review E}\ }\textbf {\bibinfo {volume} {99}},\ \href {https://doi.org/10.1103/physreve.99.032123} {10.1103/physreve.99.032123} (\bibinfo {year} {2019})\BibitemShut {NoStop}%
\bibitem [{\citenamefont {Ku\'smierz}\ and\ \citenamefont {Gudowska-Nowak}(2015)}]{Kumierz2015}%
  \BibitemOpen
  \bibfield  {author} {\bibinfo {author} {\bibfnamefont {L.}~\bibnamefont {Ku\'smierz}}\ and\ \bibinfo {author} {\bibfnamefont {E.}~\bibnamefont {Gudowska-Nowak}},\ }\bibfield  {journal} {\bibinfo  {journal} {Physical Review E}\ }\textbf {\bibinfo {volume} {92}},\ \href {https://doi.org/10.1103/physreve.92.052127} {10.1103/physreve.92.052127} (\bibinfo {year} {2015})\BibitemShut {NoStop}%
\bibitem [{\citenamefont {Pal}\ and\ \citenamefont {Reuveni}(2017)}]{Pal2017}%
  \BibitemOpen
  \bibfield  {author} {\bibinfo {author} {\bibfnamefont {A.}~\bibnamefont {Pal}}\ and\ \bibinfo {author} {\bibfnamefont {S.}~\bibnamefont {Reuveni}},\ }\bibfield  {journal} {\bibinfo  {journal} {Physical Review Letters}\ }\textbf {\bibinfo {volume} {118}},\ \href {https://doi.org/10.1103/physrevlett.118.030603} {10.1103/physrevlett.118.030603} (\bibinfo {year} {2017})\BibitemShut {NoStop}%
\bibitem [{\citenamefont {Das}\ and\ \citenamefont {Giuggioli}(2022)}]{Das2022}%
  \BibitemOpen
  \bibfield  {author} {\bibinfo {author} {\bibfnamefont {D.}~\bibnamefont {Das}}\ and\ \bibinfo {author} {\bibfnamefont {L.}~\bibnamefont {Giuggioli}},\ }\href {https://doi.org/10.1088/1751-8121/ac9765} {\bibfield  {journal} {\bibinfo  {journal} {Journal of Physics A: Mathematical and Theoretical}\ }\textbf {\bibinfo {volume} {55}},\ \bibinfo {pages} {424004} (\bibinfo {year} {2022})}\BibitemShut {NoStop}%
\bibitem [{\citenamefont {Ramoso}\ \emph {et~al.}(2020)\citenamefont {Ramoso}, \citenamefont {Magalang}, \citenamefont {S\'anchez-Taltavull}, \citenamefont {Esguerra},\ and\ \citenamefont {Rold\'an}}]{ramoso_stochastic_2020}%
  \BibitemOpen
  \bibfield  {author} {\bibinfo {author} {\bibfnamefont {A.}~\bibnamefont {Ramoso}}, \bibinfo {author} {\bibfnamefont {J.}~\bibnamefont {Magalang}}, \bibinfo {author} {\bibfnamefont {D.}~\bibnamefont {S\'anchez-Taltavull}}, \bibinfo {author} {\bibfnamefont {J.}~\bibnamefont {Esguerra}},\ and\ \bibinfo {author} {\bibfnamefont {E.}~\bibnamefont {Rold\'an}},\ }\href {https://doi.org/10.1209/0295-5075/132/50003} {\bibfield  {journal} {\bibinfo  {journal} {EPL (Europhysics Letters)}\ }\textbf {\bibinfo {volume} {132}},\ \bibinfo {pages} {50003} (\bibinfo {year} {2020})}\BibitemShut {NoStop}%
\bibitem [{\citenamefont {Dy}\ and\ \citenamefont {Esguerra}(2013)}]{dyesguerra}%
  \BibitemOpen
  \bibfield  {author} {\bibinfo {author} {\bibfnamefont {D.~L.}\ \bibnamefont {Dy}}\ and\ \bibinfo {author} {\bibfnamefont {J.~P.}\ \bibnamefont {Esguerra}},\ }\href {https://doi.org/10.1103/PhysRevE.88.012121} {\bibfield  {journal} {\bibinfo  {journal} {Phys. Rev. E}\ }\textbf {\bibinfo {volume} {88}},\ \bibinfo {pages} {012121} (\bibinfo {year} {2013})}\BibitemShut {NoStop}%
\bibitem [{\citenamefont {Gardiner}(2008)}]{Gardiner2008-re}%
  \BibitemOpen
  \bibfield  {author} {\bibinfo {author} {\bibfnamefont {C.}~\bibnamefont {Gardiner}},\ }in\ \href@noop {} {\emph {\bibinfo {booktitle} {Stochastic Methods}}},\ \bibinfo {series and number} {Springer series in synergetics}\ (\bibinfo  {publisher} {Springer Berlin Heidelberg},\ \bibinfo {address} {Berlin, Heidelberg},\ \bibinfo {year} {2008})\ pp.\ \bibinfo {pages} {1--22}\BibitemShut {NoStop}%
\bibitem [{\citenamefont {Giuggioli}\ \emph {et~al.}(2023)\citenamefont {Giuggioli}, \citenamefont {Sarvaharman}, \citenamefont {Das}, \citenamefont {Marris},\ and\ \citenamefont {Kay}}]{giuggioli2023}%
  \BibitemOpen
  \bibfield  {author} {\bibinfo {author} {\bibfnamefont {L.}~\bibnamefont {Giuggioli}}, \bibinfo {author} {\bibfnamefont {S.}~\bibnamefont {Sarvaharman}}, \bibinfo {author} {\bibfnamefont {D.}~\bibnamefont {Das}}, \bibinfo {author} {\bibfnamefont {D.}~\bibnamefont {Marris}},\ and\ \bibinfo {author} {\bibfnamefont {T.}~\bibnamefont {Kay}},\ }\href {https://doi.org/10.48550/ARXIV.2311.00464} {\bibinfo {title} {Multi-target search in bounded and heterogeneous environments: a lattice random walk perspective}} (\bibinfo {year} {2023})\BibitemShut {NoStop}%
\bibitem [{\citenamefont {Kampen}(2007)}]{vankampen2007}%
  \BibitemOpen
  \bibfield  {author} {\bibinfo {author} {\bibfnamefont {N.~G.~V.}\ \bibnamefont {Kampen}},\ }\href {https://doi.org/10.1016/b978-0-444-52965-7.x5000-4} {\emph {\bibinfo {title} {Stochastic Processes in Physics and Chemistry}}}\ (\bibinfo  {publisher} {Elsevier},\ \bibinfo {year} {2007})\BibitemShut {NoStop}%
\bibitem [{\citenamefont {Gillespie}(1977)}]{Gillespie1977}%
  \BibitemOpen
  \bibfield  {author} {\bibinfo {author} {\bibfnamefont {D.~T.}\ \bibnamefont {Gillespie}},\ }\href {https://doi.org/10.1021/j100540a008} {\bibfield  {journal} {\bibinfo  {journal} {The Journal of Physical Chemistry}\ }\textbf {\bibinfo {volume} {81}},\ \bibinfo {pages} {2340–2361} (\bibinfo {year} {1977})}\BibitemShut {NoStop}%
\bibitem [{\citenamefont {Gillespie}(1980)}]{Gillespie1980}%
  \BibitemOpen
  \bibfield  {author} {\bibinfo {author} {\bibfnamefont {D.~T.}\ \bibnamefont {Gillespie}},\ }\href {https://doi.org/10.1063/1.439029} {\bibfield  {journal} {\bibinfo  {journal} {The Journal of Chemical Physics}\ }\textbf {\bibinfo {volume} {72}},\ \bibinfo {pages} {5363} (\bibinfo {year} {1980})}\BibitemShut {NoStop}%
\bibitem [{\citenamefont {Harunari}\ \emph {et~al.}(2022)\citenamefont {Harunari}, \citenamefont {Dutta}, \citenamefont {Polettini},\ and\ \citenamefont {Rold\'an}}]{Harunari2022}%
  \BibitemOpen
  \bibfield  {author} {\bibinfo {author} {\bibfnamefont {P.~E.}\ \bibnamefont {Harunari}}, \bibinfo {author} {\bibfnamefont {A.}~\bibnamefont {Dutta}}, \bibinfo {author} {\bibfnamefont {M.}~\bibnamefont {Polettini}},\ and\ \bibinfo {author} {\bibfnamefont {E.}~\bibnamefont {Rold\'an}},\ }\bibfield  {journal} {\bibinfo  {journal} {Physical Review X}\ }\textbf {\bibinfo {volume} {12}},\ \href {https://doi.org/10.1103/physrevx.12.041026} {10.1103/physrevx.12.041026} (\bibinfo {year} {2022})\BibitemShut {NoStop}%
\bibitem [{\citenamefont {van~der Meer}\ \emph {et~al.}(2022)\citenamefont {van~der Meer}, \citenamefont {Ertel},\ and\ \citenamefont {Seifert}}]{vanderMeer2022}%
  \BibitemOpen
  \bibfield  {author} {\bibinfo {author} {\bibfnamefont {J.}~\bibnamefont {van~der Meer}}, \bibinfo {author} {\bibfnamefont {B.}~\bibnamefont {Ertel}},\ and\ \bibinfo {author} {\bibfnamefont {U.}~\bibnamefont {Seifert}},\ }\bibfield  {journal} {\bibinfo  {journal} {Physical Review X}\ }\textbf {\bibinfo {volume} {12}},\ \href {https://doi.org/10.1103/physrevx.12.031025} {10.1103/physrevx.12.031025} (\bibinfo {year} {2022})\BibitemShut {NoStop}%
\bibitem [{\citenamefont {Sekimoto}(2021)}]{sekimoto2021derivation}%
  \BibitemOpen
  \bibfield  {author} {\bibinfo {author} {\bibfnamefont {K.}~\bibnamefont {Sekimoto}},\ }\href@noop {} {\bibfield  {journal} {\bibinfo  {journal} {arXiv preprint arXiv:2110.02216}\ } (\bibinfo {year} {2021})}\BibitemShut {NoStop}%
\bibitem [{\citenamefont {Roemhild}\ \emph {et~al.}(2022)\citenamefont {Roemhild}, \citenamefont {Bollenbach},\ and\ \citenamefont {Andersson}}]{Roemhild2022}%
  \BibitemOpen
  \bibfield  {author} {\bibinfo {author} {\bibfnamefont {R.}~\bibnamefont {Roemhild}}, \bibinfo {author} {\bibfnamefont {T.}~\bibnamefont {Bollenbach}},\ and\ \bibinfo {author} {\bibfnamefont {D.~I.}\ \bibnamefont {Andersson}},\ }\href {https://doi.org/10.1038/s41579-022-00700-5} {\bibfield  {journal} {\bibinfo  {journal} {Nature Reviews Microbiology}\ }\textbf {\bibinfo {volume} {20}},\ \bibinfo {pages} {478–490} (\bibinfo {year} {2022})}\BibitemShut {NoStop}%
\bibitem [{\citenamefont {Sunil}\ \emph {et~al.}(2023)\citenamefont {Sunil}, \citenamefont {Blythe}, \citenamefont {Evans},\ and\ \citenamefont {Majumdar}}]{Sunil2023}%
  \BibitemOpen
  \bibfield  {author} {\bibinfo {author} {\bibfnamefont {J.~C.}\ \bibnamefont {Sunil}}, \bibinfo {author} {\bibfnamefont {R.~A.}\ \bibnamefont {Blythe}}, \bibinfo {author} {\bibfnamefont {M.~R.}\ \bibnamefont {Evans}},\ and\ \bibinfo {author} {\bibfnamefont {S.~N.}\ \bibnamefont {Majumdar}},\ }\href {https://doi.org/10.1088/1751-8121/acf3bb} {\bibfield  {journal} {\bibinfo  {journal} {Journal of Physics A: Mathematical and Theoretical}\ }\textbf {\bibinfo {volume} {56}},\ \bibinfo {pages} {395001} (\bibinfo {year} {2023})}\BibitemShut {NoStop}%
\bibitem [{\citenamefont {De~Bruyne}\ and\ \citenamefont {Mori}(2023)}]{debruyne2023}%
  \BibitemOpen
  \bibfield  {author} {\bibinfo {author} {\bibfnamefont {B.}~\bibnamefont {De~Bruyne}}\ and\ \bibinfo {author} {\bibfnamefont {F.}~\bibnamefont {Mori}},\ }\href {https://doi.org/10.1103/PhysRevResearch.5.013122} {\bibfield  {journal} {\bibinfo  {journal} {Phys. Rev. Res.}\ }\textbf {\bibinfo {volume} {5}},\ \bibinfo {pages} {013122} (\bibinfo {year} {2023})}\BibitemShut {NoStop}%
\bibitem [{\citenamefont {Baral}\ \emph {et~al.}(2019)\citenamefont {Baral}, \citenamefont {Raja}, \citenamefont {Sen},\ and\ \citenamefont {Dixit}}]{Baral2019}%
  \BibitemOpen
  \bibfield  {author} {\bibinfo {author} {\bibfnamefont {S.}~\bibnamefont {Baral}}, \bibinfo {author} {\bibfnamefont {R.}~\bibnamefont {Raja}}, \bibinfo {author} {\bibfnamefont {P.}~\bibnamefont {Sen}},\ and\ \bibinfo {author} {\bibfnamefont {N.~M.}\ \bibnamefont {Dixit}},\ }\bibfield  {journal} {\bibinfo  {journal} {WIREs Systems Biology and Medicine}\ }\textbf {\bibinfo {volume} {11}},\ \href {https://doi.org/10.1002/wsbm.1446} {10.1002/wsbm.1446} (\bibinfo {year} {2019})\BibitemShut {NoStop}%
\bibitem [{\citenamefont {Prosperi}\ \emph {et~al.}(2008)\citenamefont {Prosperi}, \citenamefont {D'Autilia}, \citenamefont {Incardona}, \citenamefont {De~Luca}, \citenamefont {Zazzi},\ and\ \citenamefont {Ulivi}}]{prosperi2008}%
  \BibitemOpen
  \bibfield  {author} {\bibinfo {author} {\bibfnamefont {M.~C.~F.}\ \bibnamefont {Prosperi}}, \bibinfo {author} {\bibfnamefont {R.}~\bibnamefont {D'Autilia}}, \bibinfo {author} {\bibfnamefont {F.}~\bibnamefont {Incardona}}, \bibinfo {author} {\bibfnamefont {A.}~\bibnamefont {De~Luca}}, \bibinfo {author} {\bibfnamefont {M.}~\bibnamefont {Zazzi}},\ and\ \bibinfo {author} {\bibfnamefont {G.}~\bibnamefont {Ulivi}},\ }\href {https://doi.org/10.1093/bioinformatics/btn568} {\bibfield  {journal} {\bibinfo  {journal} {Bioinformatics}\ }\textbf {\bibinfo {volume} {25}},\ \bibinfo {pages} {1040} (\bibinfo {year} {2008})},\ \Eprint {https://arxiv.org/abs/https://academic.oup.com/bioinformatics/article-pdf/25/8/1040/50287497/bioinformatics\_25\_8\_1040.pdf} {https://academic.oup.com/bioinformatics/article-pdf/25/8/1040/50287497/bioinformatics\_25\_8\_1040.pdf} \BibitemShut {NoStop}%
\bibitem [{\citenamefont {Nicoloff}\ \emph {et~al.}(2019)\citenamefont {Nicoloff}, \citenamefont {Hjort}, \citenamefont {Levin},\ and\ \citenamefont {Andersson}}]{Nicoloff2019}%
  \BibitemOpen
  \bibfield  {author} {\bibinfo {author} {\bibfnamefont {H.}~\bibnamefont {Nicoloff}}, \bibinfo {author} {\bibfnamefont {K.}~\bibnamefont {Hjort}}, \bibinfo {author} {\bibfnamefont {B.~R.}\ \bibnamefont {Levin}},\ and\ \bibinfo {author} {\bibfnamefont {D.~I.}\ \bibnamefont {Andersson}},\ }\href {https://doi.org/10.1038/s41564-018-0342-0} {\bibfield  {journal} {\bibinfo  {journal} {Nature Microbiology}\ }\textbf {\bibinfo {volume} {4}},\ \bibinfo {pages} {504–514} (\bibinfo {year} {2019})}\BibitemShut {NoStop}%
\bibitem [{\citenamefont {Roemhild}\ and\ \citenamefont {Schulenburg}(2019)}]{Roemhild2019}%
  \BibitemOpen
  \bibfield  {author} {\bibinfo {author} {\bibfnamefont {R.}~\bibnamefont {Roemhild}}\ and\ \bibinfo {author} {\bibfnamefont {H.}~\bibnamefont {Schulenburg}},\ }\href {https://doi.org/10.1093/emph/eoz008} {\bibfield  {journal} {\bibinfo  {journal} {Evolution, Medicine, and Public Health}\ }\textbf {\bibinfo {volume} {2019}},\ \bibinfo {pages} {37–45} (\bibinfo {year} {2019})}\BibitemShut {NoStop}%
\bibitem [{\citenamefont {Rosenkilde}\ \emph {et~al.}(2019)\citenamefont {Rosenkilde}, \citenamefont {Munck}, \citenamefont {Porse}, \citenamefont {Linkevicius}, \citenamefont {Andersson},\ and\ \citenamefont {Sommer}}]{Rosenkilde2019}%
  \BibitemOpen
  \bibfield  {author} {\bibinfo {author} {\bibfnamefont {C.~E.~H.}\ \bibnamefont {Rosenkilde}}, \bibinfo {author} {\bibfnamefont {C.}~\bibnamefont {Munck}}, \bibinfo {author} {\bibfnamefont {A.}~\bibnamefont {Porse}}, \bibinfo {author} {\bibfnamefont {M.}~\bibnamefont {Linkevicius}}, \bibinfo {author} {\bibfnamefont {D.~I.}\ \bibnamefont {Andersson}},\ and\ \bibinfo {author} {\bibfnamefont {M.~O.~A.}\ \bibnamefont {Sommer}},\ }\bibfield  {journal} {\bibinfo  {journal} {Nature Communications}\ }\textbf {\bibinfo {volume} {10}},\ \href {https://doi.org/10.1038/s41467-019-08529-y} {10.1038/s41467-019-08529-y} (\bibinfo {year} {2019})\BibitemShut {NoStop}%
\bibitem [{\citenamefont {Imamovic}\ and\ \citenamefont {Sommer}(2013)}]{Imamovic2013}%
  \BibitemOpen
  \bibfield  {author} {\bibinfo {author} {\bibfnamefont {L.}~\bibnamefont {Imamovic}}\ and\ \bibinfo {author} {\bibfnamefont {M.~O.~A.}\ \bibnamefont {Sommer}},\ }\bibfield  {journal} {\bibinfo  {journal} {Science Translational Medicine}\ }\textbf {\bibinfo {volume} {5}},\ \href {https://doi.org/10.1126/scitranslmed.3006609} {10.1126/scitranslmed.3006609} (\bibinfo {year} {2013})\BibitemShut {NoStop}%
\bibitem [{\citenamefont {Batra}\ \emph {et~al.}(2021)\citenamefont {Batra}, \citenamefont {Roemhild}, \citenamefont {Rousseau}, \citenamefont {Franzenburg}, \citenamefont {Niemann},\ and\ \citenamefont {Schulenburg}}]{Batra2021}%
  \BibitemOpen
  \bibfield  {author} {\bibinfo {author} {\bibfnamefont {A.}~\bibnamefont {Batra}}, \bibinfo {author} {\bibfnamefont {R.}~\bibnamefont {Roemhild}}, \bibinfo {author} {\bibfnamefont {E.}~\bibnamefont {Rousseau}}, \bibinfo {author} {\bibfnamefont {S.}~\bibnamefont {Franzenburg}}, \bibinfo {author} {\bibfnamefont {S.}~\bibnamefont {Niemann}},\ and\ \bibinfo {author} {\bibfnamefont {H.}~\bibnamefont {Schulenburg}},\ }\bibfield  {journal} {\bibinfo  {journal} {eLife}\ }\textbf {\bibinfo {volume} {10}},\ \href {https://doi.org/10.7554/elife.68876} {10.7554/elife.68876} (\bibinfo {year} {2021})\BibitemShut {NoStop}%
\bibitem [{\citenamefont {Oette}\ \emph {et~al.}(2012)\citenamefont {Oette}, \citenamefont {Schülter}, \citenamefont {Rosen-Zvi}, \citenamefont {Peres}, \citenamefont {Zazzi}, \citenamefont {Sönnerborg}, \citenamefont {Struck}, \citenamefont {Altmann}, \citenamefont {Kaiser},\ and\ \citenamefont {the EuResist Network Study~Group}}]{oette2012}%
  \BibitemOpen
  \bibfield  {author} {\bibinfo {author} {\bibfnamefont {M.}~\bibnamefont {Oette}}, \bibinfo {author} {\bibfnamefont {E.}~\bibnamefont {Schülter}}, \bibinfo {author} {\bibfnamefont {M.}~\bibnamefont {Rosen-Zvi}}, \bibinfo {author} {\bibfnamefont {Y.}~\bibnamefont {Peres}}, \bibinfo {author} {\bibfnamefont {M.}~\bibnamefont {Zazzi}}, \bibinfo {author} {\bibfnamefont {A.}~\bibnamefont {Sönnerborg}}, \bibinfo {author} {\bibfnamefont {D.}~\bibnamefont {Struck}}, \bibinfo {author} {\bibfnamefont {A.}~\bibnamefont {Altmann}}, \bibinfo {author} {\bibfnamefont {R.}~\bibnamefont {Kaiser}},\ and\ \bibinfo {author} {\bibnamefont {the EuResist Network Study~Group}},\ }\href {https://doi.org/10.1159/000332018} {\bibfield  {journal} {\bibinfo  {journal} {Intervirology}\ }\textbf {\bibinfo {volume} {55}},\ \bibinfo {pages} {160} (\bibinfo {year} {2012})},\ \Eprint {https://arxiv.org/abs/https://karger.com/int/article-pdf/55/2/160/2942682/000332018.pdf} {https://karger.com/int/article-pdf/55/2/160/2942682/000332018.pdf}
  \BibitemShut {NoStop}%
\bibitem [{\citenamefont {Sochman}\ and\ \citenamefont {Podzimkova}(2005)}]{Sochman2005}%
  \BibitemOpen
  \bibfield  {author} {\bibinfo {author} {\bibfnamefont {J.}~\bibnamefont {Sochman}}\ and\ \bibinfo {author} {\bibfnamefont {M.}~\bibnamefont {Podzimkova}},\ }\href {https://doi.org/10.1016/j.ijcard.2003.11.019} {\bibfield  {journal} {\bibinfo  {journal} {International Journal of Cardiology}\ }\textbf {\bibinfo {volume} {99}},\ \bibinfo {pages} {145–146} (\bibinfo {year} {2005})}\BibitemShut {NoStop}%
\bibitem [{\citenamefont {Avanceña}\ and\ \citenamefont {Hutton}(2020)}]{AVANCENA20201509}%
  \BibitemOpen
  \bibfield  {author} {\bibinfo {author} {\bibfnamefont {A.~L.}\ \bibnamefont {Avanceña}}\ and\ \bibinfo {author} {\bibfnamefont {D.~W.}\ \bibnamefont {Hutton}},\ }\href {https://doi.org/https://doi.org/10.1016/j.jval.2020.08.001} {\bibfield  {journal} {\bibinfo  {journal} {Value in Health}\ }\textbf {\bibinfo {volume} {23}},\ \bibinfo {pages} {1509} (\bibinfo {year} {2020})}\BibitemShut {NoStop}%
\bibitem [{\citenamefont {Brandeau}\ and\ \citenamefont {Zaric}(2008)}]{Brandeau2008}%
  \BibitemOpen
  \bibfield  {author} {\bibinfo {author} {\bibfnamefont {M.~L.}\ \bibnamefont {Brandeau}}\ and\ \bibinfo {author} {\bibfnamefont {G.~S.}\ \bibnamefont {Zaric}},\ }\href {https://doi.org/10.1007/s10729-008-9074-7} {\bibfield  {journal} {\bibinfo  {journal} {Health Care Management Science}\ }\textbf {\bibinfo {volume} {12}},\ \bibinfo {pages} {27–37} (\bibinfo {year} {2008})}\BibitemShut {NoStop}%
\bibitem [{\citenamefont {Duwal}\ \emph {et~al.}(2015)\citenamefont {Duwal}, \citenamefont {Winkelmann}, \citenamefont {Sch\"{u}tte},\ and\ \citenamefont {von Kleist}}]{Duwal2015}%
  \BibitemOpen
  \bibfield  {author} {\bibinfo {author} {\bibfnamefont {S.}~\bibnamefont {Duwal}}, \bibinfo {author} {\bibfnamefont {S.}~\bibnamefont {Winkelmann}}, \bibinfo {author} {\bibfnamefont {C.}~\bibnamefont {Sch\"{u}tte}},\ and\ \bibinfo {author} {\bibfnamefont {M.}~\bibnamefont {von Kleist}},\ }\href {https://doi.org/10.1371/journal.pcbi.1004200} {\bibfield  {journal} {\bibinfo  {journal} {PLOS Computational Biology}\ }\textbf {\bibinfo {volume} {11}},\ \bibinfo {pages} {e1004200} (\bibinfo {year} {2015})}\BibitemShut {NoStop}%
\bibitem [{\citenamefont {Greco}\ \emph {et~al.}(1995)\citenamefont {Greco}, \citenamefont {Bravo},\ and\ \citenamefont {Parsons}}]{Greco1995-ei}%
  \BibitemOpen
  \bibfield  {author} {\bibinfo {author} {\bibfnamefont {W.~R.}\ \bibnamefont {Greco}}, \bibinfo {author} {\bibfnamefont {G.}~\bibnamefont {Bravo}},\ and\ \bibinfo {author} {\bibfnamefont {J.~C.}\ \bibnamefont {Parsons}},\ }\href@noop {} {\bibfield  {journal} {\bibinfo  {journal} {Pharmacol. Rev.}\ }\textbf {\bibinfo {volume} {47}},\ \bibinfo {pages} {331} (\bibinfo {year} {1995})}\BibitemShut {NoStop}%
\bibitem [{\citenamefont {BLISS}(1939)}]{BLISS1939}%
  \BibitemOpen
  \bibfield  {author} {\bibinfo {author} {\bibfnamefont {C.~I.}\ \bibnamefont {BLISS}},\ }\href {https://doi.org/10.1111/j.1744-7348.1939.tb06990.x} {\bibfield  {journal} {\bibinfo  {journal} {Annals of Applied Biology}\ }\textbf {\bibinfo {volume} {26}},\ \bibinfo {pages} {585–615} (\bibinfo {year} {1939})}\BibitemShut {NoStop}%
\bibitem [{\citenamefont {Chen}\ \emph {et~al.}(2024)\citenamefont {Chen}, \citenamefont {Liu}, \citenamefont {Du}, \citenamefont {Wee}, \citenamefont {Wang}, \citenamefont {Chen}, \citenamefont {Shen},\ and\ \citenamefont {Wei}}]{chen2024drug}%
  \BibitemOpen
  \bibfield  {author} {\bibinfo {author} {\bibfnamefont {D.}~\bibnamefont {Chen}}, \bibinfo {author} {\bibfnamefont {G.}~\bibnamefont {Liu}}, \bibinfo {author} {\bibfnamefont {H.}~\bibnamefont {Du}}, \bibinfo {author} {\bibfnamefont {J.}~\bibnamefont {Wee}}, \bibinfo {author} {\bibfnamefont {R.}~\bibnamefont {Wang}}, \bibinfo {author} {\bibfnamefont {J.}~\bibnamefont {Chen}}, \bibinfo {author} {\bibfnamefont {J.}~\bibnamefont {Shen}},\ and\ \bibinfo {author} {\bibfnamefont {G.-W.}\ \bibnamefont {Wei}},\ }\href@noop {} {\bibfield  {journal} {\bibinfo  {journal} {ArXiv}\ } (\bibinfo {year} {2024})}\BibitemShut {NoStop}%
\bibitem [{\citenamefont {Lamirande}\ \emph {et~al.}(2024)\citenamefont {Lamirande}, \citenamefont {Gaffney}, \citenamefont {Gertz}, \citenamefont {Maini}, \citenamefont {Crawshaw},\ and\ \citenamefont {Caruso}}]{lamirande2024first}%
  \BibitemOpen
  \bibfield  {author} {\bibinfo {author} {\bibfnamefont {P.}~\bibnamefont {Lamirande}}, \bibinfo {author} {\bibfnamefont {E.~A.}\ \bibnamefont {Gaffney}}, \bibinfo {author} {\bibfnamefont {M.}~\bibnamefont {Gertz}}, \bibinfo {author} {\bibfnamefont {P.~K.}\ \bibnamefont {Maini}}, \bibinfo {author} {\bibfnamefont {J.~R.}\ \bibnamefont {Crawshaw}},\ and\ \bibinfo {author} {\bibfnamefont {A.}~\bibnamefont {Caruso}},\ }\href@noop {} {\bibfield  {journal} {\bibinfo  {journal} {arXiv preprint arXiv:2404.04086}\ } (\bibinfo {year} {2024})}\BibitemShut {NoStop}%
\bibitem [{\citenamefont {Lebovitz}\ and\ \citenamefont {Banerji}(2004)}]{Lebovitz2004}%
  \BibitemOpen
  \bibfield  {author} {\bibinfo {author} {\bibfnamefont {H.~E.}\ \bibnamefont {Lebovitz}}\ and\ \bibinfo {author} {\bibfnamefont {M.~A.}\ \bibnamefont {Banerji}},\ }\href {https://doi.org/10.1016/j.ejphar.2004.02.051} {\bibfield  {journal} {\bibinfo  {journal} {European Journal of Pharmacology}\ }\textbf {\bibinfo {volume} {490}},\ \bibinfo {pages} {135–146} (\bibinfo {year} {2004})}\BibitemShut {NoStop}%
\bibitem [{\citenamefont {Braido}\ \emph {et~al.}(2015)\citenamefont {Braido}, \citenamefont {Lavorini}, \citenamefont {Blasi}, \citenamefont {Baiardini},\ and\ \citenamefont {Canonica}}]{Braido2015}%
  \BibitemOpen
  \bibfield  {author} {\bibinfo {author} {\bibfnamefont {F.}~\bibnamefont {Braido}}, \bibinfo {author} {\bibfnamefont {F.}~\bibnamefont {Lavorini}}, \bibinfo {author} {\bibfnamefont {F.}~\bibnamefont {Blasi}}, \bibinfo {author} {\bibfnamefont {I.}~\bibnamefont {Baiardini}},\ and\ \bibinfo {author} {\bibfnamefont {G.~W.}\ \bibnamefont {Canonica}},\ }\href {https://doi.org/10.2147/copd.s79635} {\bibfield  {journal} {\bibinfo  {journal} {International Journal of Chronic Obstructive Pulmonary Disease}\ ,\ \bibinfo {pages} {2601}} (\bibinfo {year} {2015})}\BibitemShut {NoStop}%
\bibitem [{\citenamefont {Wong}\ \emph {et~al.}(2013)\citenamefont {Wong}, \citenamefont {Tam}, \citenamefont {Cheung}, \citenamefont {Tong}, \citenamefont {Sek}, \citenamefont {John}, \citenamefont {Cheung}, \citenamefont {Yan}, \citenamefont {Yu}, \citenamefont {Leeder},\ and\ \citenamefont {Griffiths}}]{Wong2013}%
  \BibitemOpen
  \bibfield  {author} {\bibinfo {author} {\bibfnamefont {M.~C.~S.}\ \bibnamefont {Wong}}, \bibinfo {author} {\bibfnamefont {W.~W.~S.}\ \bibnamefont {Tam}}, \bibinfo {author} {\bibfnamefont {C.~S.~K.}\ \bibnamefont {Cheung}}, \bibinfo {author} {\bibfnamefont {E.~L.~H.}\ \bibnamefont {Tong}}, \bibinfo {author} {\bibfnamefont {A.~C.~H.}\ \bibnamefont {Sek}}, \bibinfo {author} {\bibfnamefont {G.}~\bibnamefont {John}}, \bibinfo {author} {\bibfnamefont {N.~T.}\ \bibnamefont {Cheung}}, \bibinfo {author} {\bibfnamefont {B.~P.~Y.}\ \bibnamefont {Yan}}, \bibinfo {author} {\bibfnamefont {C.~M.}\ \bibnamefont {Yu}}, \bibinfo {author} {\bibfnamefont {S.}~\bibnamefont {Leeder}},\ and\ \bibinfo {author} {\bibfnamefont {S.}~\bibnamefont {Griffiths}},\ }\href {https://doi.org/10.1371/journal.pone.0053625} {\bibfield  {journal} {\bibinfo  {journal} {PLoS ONE}\ }\textbf {\bibinfo {volume} {8}},\ \bibinfo {pages} {e53625} (\bibinfo {year} {2013})}\BibitemShut {NoStop}%
\bibitem [{\citenamefont {V\'aclav\'ik}\ \emph {et~al.}(2014)\citenamefont {V\'aclav\'ik}, \citenamefont {Vyso{\v c}anov\'a}, \citenamefont {Seidlerov\'a}, \citenamefont {Zaj\'i{\v c}ek}, \citenamefont {Petr\'ak}, \citenamefont {Dlask},\ and\ \citenamefont {Kr\'yza}}]{Vclavk2014}%
  \BibitemOpen
  \bibfield  {author} {\bibinfo {author} {\bibfnamefont {J.}~\bibnamefont {V\'aclav\'ik}}, \bibinfo {author} {\bibfnamefont {P.}~\bibnamefont {Vyso{\v c}anov\'a}}, \bibinfo {author} {\bibfnamefont {J.}~\bibnamefont {Seidlerov\'a}}, \bibinfo {author} {\bibfnamefont {P.}~\bibnamefont {Zaj\'i{\v c}ek}}, \bibinfo {author} {\bibfnamefont {O.}~\bibnamefont {Petr\'ak}}, \bibinfo {author} {\bibfnamefont {J.}~\bibnamefont {Dlask}},\ and\ \bibinfo {author} {\bibfnamefont {J.}~\bibnamefont {Kr\'yza}},\ }\href {https://doi.org/10.1097/md.0000000000000168} {\bibfield  {journal} {\bibinfo  {journal} {Medicine}\ }\textbf {\bibinfo {volume} {93}},\ \bibinfo {pages} {e168} (\bibinfo {year} {2014})}\BibitemShut {NoStop}%
\bibitem [{\citenamefont {Athyros}\ \emph {et~al.}(2010)\citenamefont {Athyros}, \citenamefont {Tziomalos}, \citenamefont {Karagiannis},\ and\ \citenamefont {Mikhailidis}}]{Athyros2010}%
  \BibitemOpen
  \bibfield  {author} {\bibinfo {author} {\bibfnamefont {V.~G.}\ \bibnamefont {Athyros}}, \bibinfo {author} {\bibfnamefont {K.}~\bibnamefont {Tziomalos}}, \bibinfo {author} {\bibfnamefont {A.}~\bibnamefont {Karagiannis}},\ and\ \bibinfo {author} {\bibfnamefont {D.~P.}\ \bibnamefont {Mikhailidis}},\ }\href {https://doi.org/10.1517/14656566.2010.522991} {\bibfield  {journal} {\bibinfo  {journal} {Expert Opinion on Pharmacotherapy}\ }\textbf {\bibinfo {volume} {11}},\ \bibinfo {pages} {2943–2946} (\bibinfo {year} {2010})}\BibitemShut {NoStop}%
\end{thebibliography}%

\clearpage

\appendix

\section{Steady states of the host-pathogen model}\label{AP:fixedpt}

In this section, we will describe how the quarter-circle absorbing boundary condition described in Section \ref{sec:models} and Fig. \hyperref[fig:2thdiagram]{1c} is obtained from the steady states of a host-pathogen model. The host-pathogen model considered in this paper is a model of chronic infection, specifically HIV-1 \cite{Rong2009, Callaway2002}. This model is modified to consider $N_T$ simultaneous and independent therapies, with the $i$\textsuperscript{th} having efficacy $\eta_i$. The therapy efficacies are multiplicative, following the framework for independently-acting therapies \cite{Greco1995-ei, BLISS1939, Roemhild2022}. This model tracks three classes of host cells: healthy cells $H$, latently-infecting cells $L$, and actively-infecting cells $I$:
\begin{equation}\label{eq:hostpathogen}
\begin{aligned}
        \dot{H} &= \alpha - \lambda_H H - \beta H I\prod_{i=1}^{N_T}(1-\eta_i), \\
        \hspace{-0.5cm}\dot{L} &= \epsilon\beta HI \prod_{i=1}^{N_T}(1-\eta_i)+ pL -\frac{pL^2}{K} -a_L L -\lambda_L L, \\
        \hspace{-0.5cm}\dot{I} &= (1-\epsilon)\beta HI\prod_{i=1}^{N_T}(1-\eta_i) + a_L L -\lambda_I I.
\end{aligned}
\end{equation}

The rates and parameters of the model are described as follows: Healthy cells are generated with a constant rate of $\alpha$ and die at a rate of $\lambda_H H$, where $\lambda_H$ is a death constant for healthy cells. Furthermore, these healthy cells are converted into either latently-infecting or actively infecting cells with a rate $\beta H I \prod_{i=1}^{N_T}(1-\eta_i)$, with $\beta$ as a constant of infection and $\eta$ as the therapy efficacy. The constant $\epsilon$ that scales the infection rate in $\dot{L}$ is the probability that infection will yield a latently-infected cell. Conversely, the constant $(1-\epsilon)$ scaling the infection rate in $\dot{I}$ is the probability that the infection will yield and actively-infecting cell. Latently-infecting cells proliferate with a rate $pL$, with $p$ as a constant of proliferation, limited by a carrying capacity $K$ with a rate $(pL^2)/K$, and die off with a rate $\lambda_L L$. Finally, latently-infecting cells can transform to actively-infecting cells with a rate $aL$, with $a$ as a constant of activation, and actively-infecting cells die off with a rate of $\lambda_I I$.

We assume that the host-pathogen model is at equilibrium, hence we solve for the fixed points of the model by setting each equation in \eqref{eq:hostpathogen} equal to zero and solving for $H$,$L$, and $I$, which we now denote as $\widetilde{H}$, $\widetilde{L}$, and $\widetilde{I}$. We first obtain a trivial fixed point at $\widetilde{H} = \alpha/\lambda_H, \widetilde{L} = 0, \widetilde{I} = 0$ and a non-trivial fixed point given by the following:
\begin{equation}
\begin{aligned}
    \widetilde{H} &= \dfrac{\alpha}{\lambda_H + b \widetilde{I}}, \\
    \widetilde{L} &= \dfrac{[\lambda_I b \widetilde{I} + (\lambda_I \lambda_H - (1-\epsilon) b \alpha)]\widetilde{I}}{a_L (b \widetilde{I} + \lambda_H)}, \\
    \widetilde{I} &= S + T - \frac{x_2}{3x_1}.
\end{aligned}
\label{eq:fixedptsoln}
\end{equation}

Where,
\begin{equation}
    \begin{aligned}
        S &= \sqrt[3]{R + \sqrt{Q^3 + R^2}}, \\
        T &= \sqrt[3]{R - \sqrt{Q^3 + R^2}}, \\
        Q &= \dfrac{3 x_1 x_3 - x_2^2}{ax_1^2}, \\
        R &= \dfrac{9 x_1 x_2 x_3 - 27 x_1^2 x_4 - 2x_2^2}{54 x_1 ^3}, \\
        x_1 &= -p \lambda_I^2 b^2, \\
        x_2 &= -(2p \lambda_I bc - dKa_L \lambda_I b^2), \\
        x_3 &= \epsilon b^2 \alpha K a_L^2 - pc^2 + dKa_L \lambda_H -\lambda_I b, \\
        x_4 &= \epsilon b \alpha K a_L^2 \lambda_H + dKa_L \lambda_Hc, \\
        b &= \prod_{i=1}^{N_T}( 1-\eta_i) \beta, \\
        c &= \lambda_I \lambda_H - (1-\epsilon)b \alpha, \\
        d &= p - a_L - \lambda_L.
    \end{aligned}
\end{equation}

The fixed point $\widetilde{H}$ is used to compute the diagram in Fig. \hyperref[fig:2thdiagram]{1c}, where $\widetilde{H}$ is computed for $N_T = 2$ and varying $\eta_1$ and $\eta_2$. This maps the amount of healthy cells at equilibrium to the space of therapy efficacy.

The parameter values used to generate the diagram in Fig. \hyperref[fig:2thdiagram]{1c} are as follows:

\begin{table}[!htb]
\begin{tabular}{|l|>{\raggedright}p{0.18\textwidth}|l|l|}
\hline
Parameter    & Description                                     & Value              & Unit                    \\ \hline
$H_0$          & Initial $H$ population                 & 599326             & cells/mL                \\ \hline
$L_0$          & Initial $L$ population       & 45                 & cells/mL                \\ \hline
$I_0$          & Initial $I$ population      & 11                 & cells/mL                \\ \hline
$\alpha$     & Rate of $H$ recruitment                & 6000               & cells/(mL day)          \\ \hline
$\lambda_H$  & $H$ death rate constant                    & 0.01               & 1/day \\ \hline
$\eta_i$     & Efficacy of the $i$\textsuperscript{th} therapy & varies             &                         \\ \hline
$N_T$        & Number of therapies                & varies             &                         \\ \hline
$\beta$      & Infection rate constant                         & $5 \cdot 10^{-6}$ & mL/day \\ \hline
$\epsilon$   & Fraction of infections yielding $H \to L$     & 0.01               &                         \\ \hline
$1-\epsilon$ & Fraction of infections yielding $H \to I$     & 0.99               &                         \\ \hline
$p$          & $L$ proliferation rate constant            & 0.2                & 1/day \\ \hline
$K$          & Carrying capacity of $L$ cells                  & 100                & cells/mL                \\ \hline
$a_L$        & Activation rate constant of $L$ cells           & 0.1                & 1/day \\ \hline
$\lambda_L$  & $L$ death rate constant                    & 0.01               & 1/day \\ \hline
$\lambda_I$  & $I$ death rate constant                    & 1                  & 1/day \\ \hline
\end{tabular}
\label{table:params}
\caption{Table of parameters in Fig. \hyperref[fig:2thdiagram]{1c}, adapted from a within-host model of HIV-1 \cite{Rong2009, Callaway2002, ramoso_stochastic_2020}. }
\end{table}

The values for $H_0$, $L_0$, and $I_0$ on the table have been calculated using Eq. \eqref{eq:fixedptsoln} with $N_T = 2$ and $\eta_1 = \eta_2 = 0.8$.

\section{Coupled model}\label{AP:Coupled_model}

\subsection{Model derivation}
In this section we derive Eq.~\eqref{eq:SDE_coupled} from the main text. This derivation will be performed without stochastic resetting, since the SDE with resetting is readily obtained by adding the indicator functions with the process $\chi$, as explained in Eq. \eqref{eq:SDE_general}. Consider a general time-homogeneous multidimensional SDE in the {\^I}to sense,
\begin{equation}\label{eq:general_langevin}
\begin{aligned}
    \mathrm{d}\pmb{\eta} =  \pmb{\mu}(\pmb{\eta}) \mathrm{d}t + \pmb{\sigma}(\pmb{\eta}) \mathrm{d} \pmb{W}(t).
\end{aligned}
\end{equation}
The Fokker-Planck representation of such process reads
\begin{equation}\label{eq:general_FP}
    \partial_t \rho_t(\pmb{\eta}) = -\nabla \cdot \left[\rho_t(\pmb{\eta})\pmb{\mu}(\pmb{\eta})\right]+\nabla^2 \left[\pmb{\sigma}(\pmb{\eta}) \pmb{\sigma}^\intercal(\pmb{\eta}) \rho_t\right].
\end{equation}
The above equation is invariant under orthogonal transformations of the diffusion matrix. Indeed, let $\pmb{S}$ be an orthogonal matrix such that $\pmb{S}\pmb{S}^\intercal = \pmb{I}$, then the process
\begin{equation}\label{eq:general_langevin_V2}
\begin{aligned}
    \mathrm{d}\pmb{\eta} =  \pmb{\mu}(\pmb{\eta}) \mathrm{d}t + \pmb{S}\pmb{\sigma}(\pmb{\eta}) \mathrm{d} \pmb{W}(t),
\end{aligned}
\end{equation}
is statistically indistinguishable from the process in Eq.~\eqref{eq:general_langevin} as both of them have the same associated Fokker-Planck equation \cite{Gardiner2008-re}.

Let us reduce the generality of the equations presented so far to adapt to the kind of models studied in the main text. In particular, we study isotropic and constant diffusion matrices, 
\begin{equation}
     \pmb{\sigma} = \sqrt{2D}\pmb{I}.
\end{equation}
Also, for the case of the coupled model we consider that therapies interact in such a way that the drift in the efficacy space is rotationally invariant,
\begin{equation}
     \pmb{\mu}(\pmb{\eta}) =\pmb{\eta}f(\eta),
\end{equation}
with $\eta=\sqrt{\pmb{\eta}\cdot\pmb{\eta}}$.
Thus, the resulting rotationally invariant SDE reads
\begin{equation}
\begin{aligned}
    \mathrm{d}\pmb{\eta} =  \pmb{\eta}\, f(\eta) + \sqrt{2D} \, \mathrm{d} \pmb{W}(t).
\end{aligned}
\end{equation}
The idea now is to do a change of variables to spherical coordinates with one radius $\eta$ and $N_T-1$ angles ${\phi_1,\dots,\phi_{N_T-1}}$ using Îto's rule~\cite{Gardiner2008-re,vankampen2007}. By doing so, we find that the radial drift only depends on $\eta$,
\begin{equation}
\begin{aligned}
    \sum_{i=1}^{N_T} \partial_{\eta_i} \eta \,\mathrm{d}\eta_i + D \partial_{\eta_i} ^2 \eta  \, \mathrm{d}t = \left(\eta f(\eta)+ D\frac{N_T-1}{\eta}\right)\mathrm{d}t.
\end{aligned}
\end{equation}
The radial diffusion will, in principle, depend both on the radial and angular variables. However, it can be shown that there is always an orthogonal transformation $\pmb{S}$ making the evolution for the radial component independent of the angular variables,
\begin{equation}\label{eq:langevin_invariant}
\begin{aligned}
    \mathrm{d}\eta =  \left(\eta f(\eta)+ D\frac{N_T-1}{\eta}\right)\mathrm{d}t+\sqrt{2D}\, \mathrm{d} W(t).
\end{aligned}
\end{equation}

Evolution tends to produce pathogens that are more resistant to the therapy, and consequently this means that the efficacy of the therapy will rapidly decrease. A minimal choice for the drift function $f$ that describes this mechanism is
\begin{equation}
    f(\eta) = -\frac{v}{\eta^2}.
\end{equation}
Introducing this choice for $f$ in Eq.~\eqref{eq:langevin_invariant} together with the possibility of experience stochastic resets we obtain Eq.~\eqref{eq:SDE_coupled} of the main text.

\subsection{Effect of parameters}\label{sec:effect_of_parameters}

In the main text we analyzed in detail the effect of the number of simultaneous drugs in the therapy ($N_T$) and the average frequency between therapy switches, $\tau$, as a function of the mean RDT (see Fig.~\ref{fig:effect_of_parameters} in the main text). Nevertheless, the coupled model depends on two more parameters: the diffusion constant $D$ and the constant $v$ tuning the radial drift. On the one hand, $v$ only appears in the coupled model as a subtraction to the the number of simultaneous drugs. Therefore, we can analyze changes in the drift as effective changes in the number of therapies $N_T$, which was properly analyzed. Indeed, our model with parameter choice $v=v_0+\Delta v$, $N_T=n_0$ produces the same results that realizations with $v=v_0$ and $N_T=n_0-\Delta v/D$. Thus, no new phenomenology is expected when varying $v$. On the other hand the last free parameter $D$ is also not relevant as it can be absorbed through a change of variables of the time. Thus choices on $D$ can be seen as choices of time units.  

\subsection{Mean resistance development times}

The equation for conditioned mean RDT, $\langle T  \vert \eta \rangle$, can be obtained using the backward Fokker-Planck equation formalism~\cite{Gardiner2008-re,vankampen2007}. In particular, the backward Fokker-Planck operator associated to the coupled model in Eq.~\eqref{eq:SDE_coupled} follows the equation
\begin{equation}\label{eq:AFPT}
\begin{aligned}
      D\frac{d-1}{\eta}  \frac{\mathrm{d}}{\mathrm{d}\eta} \langle T  \vert \eta \rangle+D\frac{\mathrm{d}^2}{\mathrm{d}\eta^2} \langle T  \vert \eta \rangle & \\
      +\frac{1}{\tau}\left( \langle T  \vert \eta_r \rangle-  \langle T  \vert \eta \rangle\right) &=-1.
\end{aligned}
\end{equation}
Where $d=v/D+N_T$, $\eta$ is the initial efficacy of the therapy and $\eta_r$ is the efficacy right after therapy switch. Further details on the derivation of Eq.\eqref{eq:AFPT} may be found in \cite{Evans2011, evans_diffusion_2011}. 

Using the change of variables $G(\eta)=-\langle T  \vert \eta_r \rangle+  \langle T  \vert \eta \rangle $, Eq.~\eqref{eq:AFPT} becomes
\begin{equation}\label{eq:AFPT_V2}
\begin{aligned}
    \frac{\mathrm{d}^2}{\mathrm{d}\eta^2} G(\eta)+\frac{d-1}{\eta}  \frac{\mathrm{d}}{\mathrm{d}\eta} G(\eta)+\frac{1}{\lambda^2}G(\eta)=-\frac{1}{D},
\end{aligned}
\end{equation}
with $\lambda=\sqrt{\tau D}$. Equation~\eqref{eq:AFPT_V2} has $G_p(\eta)=\tau$ as particular solution. The changes of variable $\eta=i\lambda\,x$, and $G(x)=g(x)\,x^{-\frac{d}{2}}$ transform the homogeneous part of the equation for $G$ in a Bessel equation,
\begin{equation}
    x^2 \frac{\mathrm{d}^2}{\mathrm{d}x^2} g(x)+x \frac{\mathrm{d}}{\mathrm{d}x} g(x)+(x-\beta^2)g(x)=0,
\end{equation}
with $\beta=(d/2)-1$. Therefore, the general solution for Eq.~\eqref{eq:AFPT} reads

\begin{equation}\label{eq:general_sol_MFPT}
    \langle T  \vert \eta \rangle = \eta^\beta \left[c_1 J_\beta(-i \bar{\eta}) +c_2 Y_\beta(-i \bar{\eta})\right]+\tau +c_3.
\end{equation}
With $\bar{\eta}=\eta/\lambda$. Two of the three unknown constants of Eq.~\eqref{eq:general_sol_MFPT} are determined through the boundary conditions,
\begin{align}
    \langle T  \vert \eta \rangle  (\eta_{\min}) &= 0, \\
    \frac{\mathrm{d}}{\mathrm{d}\eta}\langle T  \vert \eta  \rangle \vert _{\eta =\eta_{\max}} &= 0,
\end{align}
where $\eta_{\min}$ and $\eta_{\max}$ are the radial locations of the absorbing and reflecting boundary respectively. The third equation is provided by the self-consistent relation \begin{equation}
    \langle T  \vert \eta_r \rangle = c_3.
\end{equation}
The resulting equation for the average absorption time with initial condition $\eta$ and resetting state $\eta_r$ reads
\begin{equation}\label{eq:MFPT_general_sol}
\begin{aligned}
    \frac{\langle T  \vert \eta \rangle}{\tau} = \, \frac{J_{\beta+1}(-i\bar{\eta}_\text{max})}{Z(\eta_r)} \Biggl(&\left(\frac{\eta_r}{\eta_\text{min}}\right)^{\beta}Y_\beta(-i\bar{\eta}_\text{min}) \\
    &+\left(\frac{\eta_r}{\eta}\right)^{\beta}Y_\beta(-i\bar{\eta})\Biggr)  \\
    -\frac{Y_{\beta+1}(-i\bar{\eta}_\text{max})}{Z(\eta_r)} \Biggl(&\left(\frac{\eta_r}{\eta_\text{min}}\right)^{\beta}J_\beta(-i\bar{\eta}_\text{min}) \\
    &+\left(\frac{\eta_r}{\eta}\right)^{\beta}J_\beta(-i\bar{\eta})\Biggr) ,
\end{aligned}
\end{equation}
with
\begin{equation}
\begin{aligned}
    Z(\eta) =& \, J_{\beta+1}(-i\bar{\eta}_\text{max})Y_\beta(-i\frac{\eta}{\lambda}) \\
    &-Y_{\beta+1}(-i\bar{\eta}_\text{max})J_\beta(-i\frac{\eta}{\lambda}),
\end{aligned}
\end{equation}
and $\bar{\eta}=\eta/\lambda$, $\bar{\eta}_\text{max}=\eta_\text{max}/\lambda$, and $\bar{\eta}_\text{min}=\eta_\text{min}/\lambda$.
Eq.~\eqref{eq:AFPT_coupled} is obtained from Eq.\eqref{eq:MFPT_general_sol} fixing the resetting state to be equal to the initial condition of the process, $\eta_r=\eta$.

\section{Derivation of uncoupled model}\label{AP:uncoupled_model}

\subsection{Markov chain transition rates}\label{AP:transrates}

In this section, we will derive the transition rates $p_i$ and $q_i$ as outlined in Section \ref{subsec:uncoupled}. The transitions from one state to another can be written as a Master equation:
\begin{equation}\label{eq:master}
    \begin{aligned}
        \frac{\mathrm{d}p(\pmb{\eta},t)}{\mathrm{d}t} =& \sum_{i=1}^{N_T} \frac{1}{\tau_i} \delta (\eta_i-\eta_{i;0}) \\
        &+p_i p\left(\eta_i-\frac{1}{M}, t\right) \\
        &+ q_i p\left(\eta_i+\frac{1}{M}, t\right) \\
        &-\left(p_i+q_i+\frac{1}{\tau_i}\right) p(\eta_i,t) \, ,
    \end{aligned}
\end{equation}

where $\eta_{i;0}$ is the initial therapy efficacy of the $i$\textsuperscript{th} therapy and $\delta(\eta_i-\eta_{i;0})$ is a Dirac delta function centered at the displacement of the current efficacy from the initial efficacy $\eta_i - \eta_{i;0}$. We note that this master equation is similar in form to the master equation presented in \cite{evans_diffusion_2011}, but is generalized by considering a Markov process on a lattice.

Terms $p_i$ and $q_i$ are the Markov jump rates to adjacent states that respectively increase or decrease the therapy efficacy. Taking a Kramers-Moyal expansion to the above allows us to obtain a Fokker-Planck equation \cite{Gillespie1980, vankampen2007}. This is done by performing a second-order Taylor expansion on the terms in the master equation that are shifted by $1/M$. This expansion is performed such that the $1/M$ terms will be factored outside of the propagators, i.e. expanding $p(\eta_i+(1/M), \ldots, \eta_{N_T}, t)$ centered at $\eta_i+(1/M) = \eta_i$, $p(\eta_i-(1/M), \ldots, \eta_{N_T}, t)$ at $\eta_i-(1/M) = \eta_i$, and so on. Performing these expansions and simplifying,
\begin{equation}\label{eq:FKE}
    \begin{aligned}
        \frac{\mathrm{d}p(\pmb{\eta},t)}{\mathrm{d}t} =& \sum_{i = 1}^{N_T} \frac{1}{\tau_i} \delta (\eta_i-\eta_{1;0}) - \frac{1}{\tau_i} p(\eta_i,t) \\
        &-\dfrac{p_i-q_i}{M} \frac{\partial p(\eta_i,t)}{\partial \eta_i} \\
        &+\dfrac{p_i+q_i}{2M^2} \frac{\partial^2 p(\eta_i,t)}{\partial \eta_1^2} \\
    \end{aligned}
\end{equation}
with the following effective drift and diffusion parameters:
\begin{equation}\label{eq:vdtokj}
    v_i = \dfrac{p_i-q_i}{M}, \quad
    D_i = \dfrac{p_i+q_i}{2M^2}.
\end{equation}
Inverting these parameters,
\begin{equation}
    \begin{aligned}
    p_i &= \dfrac{1}{2}\left(2D_i M^2 + v_i M \right) \\
    q_i &= \dfrac{1}{2}\left(2D_i M^2 - v_i M \right).
    \end{aligned}
\end{equation}
We obtain the jump rates in the discrete space in terms of parameters used by the continuous space. As such, we are able to use the master equation formulation by constructing a transition matrix as discussed in the following section.

\subsection{Generating the transition matrix}\label{AP:transmatrix}

In this section, we will outline how the transition matrix for the uncoupled discrete model is generated to be used for the mean RDT formula in Eq. \eqref{eq:discmfpt}. Sums of transition rates $p_i$ and $q_i$, and therapy switching rate $1/\tau_i$ occupy the elements of the transition matrix $\matr{W}$ in Eq. \eqref{eq:gridmastereqn}. In practice, to make the transition matrix $\matr{W}$, it is simpler to start with the transition matrix from a complete lattice graph, then add and remove nodes and edges as needed until we end up with the desired graph seen in Fig. \hyperref[fig:multi-drug_model]{2b}.

Recall that the therapy efficacy vector $\pmb{\eta}$ may be viewed as an ordered coordinate on a $M^{N_T} \times M^{N_T}$ lattice of states, where $M$ is the number of therapy efficacy states for each therapy. Consider $N_T = 2$ therapies and two states $\pmb{\eta}_a = (\eta_1^{(a)}, \eta_2^{(a)})$ and $\pmb{\eta}_b = (\eta_1^{(b)}, \eta_2^{(b)})$. 

We first form the transition matrix for the complete lattice graph $\matr{M}$ with matrix elements $M_{ba}$ denoting a transition from state $\pmb{\eta}_a$ to state $\pmb{\eta}_b$. Element $M_{ba}$ is equal to a transition rate $p$ or $q$ if states $\pmb{\eta}_a$ and $\pmb{\eta}_b$ are adjacent to each other on the lattice. This element is equal to transition rate $p$ if the transition $\pmb{\eta}_a \to \pmb{\eta}_b$ increases exactly one therapy in $\pmb{\eta}_a$ by $1/M$. On the other hand, this element is equal to $q$ if the transition $ \pmb{\eta}_a \to \pmb{\eta}_b$ decreases exactly one therapy in $\pmb{\eta}_a$ by $1/M$. Diagonal elements of $\matr{M}$ are $M_{bb} = -\sum_{a \neq b} M_{ba}$ for each $b = 1, \ldots, M^2$. Any other matrix elements are zero.

Next, we form the transition matrix $\matr{R}$ that captures the behavior of therapy switching that returns the state back to its initial position. Let $\pmb{\eta}_0 = (\eta_1^{(0)}, \eta_2^{(0)})$ be the initial state and $\pmb{\eta}_b$ be any state that is not an initial state on a lattice generated by $\matr{M}$. Transition matrix $\matr{R}$ has elements $R_{ba}$ that is equal to $1/\tau_1$ for all states $\pmb{\eta}_b = (\eta_1^{(0)}, \eta_2^{(b)})$ and $1/\tau_2$ for all states $\pmb{\eta}_b = (\eta_1^{(b)}, \eta_2^{(0)})$, for $b = 1 \ldots, M^2$. Diagonal elements of $\matr{R}$ are $R_{bb} = -\sum_{a \neq b} R_{ba}$ for each $b = 1, \ldots, M^2$. Any other matrix elements are zero.

We then form the transition matrix for partial absorption along the axes of the lattice $\matr{P}$, signifying the failure of a single therapy with matrix elements $P_{ba}$. Element $P_{ba}$ is equal to transition rate $p$ if two states $\pmb{\eta}_a$ and $\pmb{\eta}_b$ are equal to $(0,\eta_2^{(a)})$ and $(1/M,\eta_2^{(b)})$, respectively or $(\eta_1^{(a)}, 0)$ and $(\eta_1^{(b)},1/M)$ respectively. Here, state $\pmb{\eta}_a$ are the states along the axis of the lattice, and $\pmb{\eta}_b$ is the immediate adjacent state to $\pmb{\eta}_a$ that is not another state on the axis, and the transitions refer to transitions from $\pmb{\eta}_a \to \pmb{\eta}_b$. Similar to the previous two matrices, diagonal elements are $P_{bb} = -\sum_{a \neq b} P_{ba}$ for each $b = 1, \ldots, M^2$ and any other matrix elements are zero.

Finally, the transition matrix for complete absorption $\matr{C}$ is formed by taking all states that fall below the absorbing condition $\sqrt{\sum_{i=1}^{N_T} \eta_i^2} \leq \eta_{\min}$ for $\eta_i \in \pmb{\eta}_b$. Matrix elements $C_{ba}$ are equal to $M_{ba}$ if state $\pmb{\eta}_b$ follow this condition.

These four matrices are added together to form the complete transition matrix $\matr{W}$ for $N_T = 2$,
\begin{equation}
\matr{W} = \matr{M}+\matr{R}-\matr{C}-\matr{P}
\end{equation}
With this transition matrix, we may use expressions that are available to Markov chains \cite{Harunari2022, vanderMeer2022, sekimoto2021derivation} in computing for the RDT statistics, as discussed further in Section \ref{subsec:uncoupled}.

\section{Unconditioned mean resistance development time} \label{ap:uncondRDT}

In this section, we will derive the mean RDT for the coupled continuous model with a random therapy efficacy after a switch. In terms of stochastic resetting, the resetting position after a reset is randomly and uniformly distributed in the state space. The unconditioned mean RDT is obtained from its conditioned version through marginalization,
\begin{equation}\label{eq:marginalization_RDT}
    \langle  T  \rangle = \int d\pmb{\eta} \,\rho\left(\pmb{\eta}\right)  \, \langle T  \vert \pmb{\eta}  \rangle,
\end{equation}
where $\rho\left(\pmb{\eta}\right)$ is the distribution of therapy efficacy right after drug switch. When the distribution $\rho\left(\pmb{\eta}\right)$ is uniform, Eq.\eqref{eq:marginalization_RDT} becomes the integral of $\langle T  \vert \pmb{\eta}  \rangle$ over all possible values of the therapy efficacy after therapy switch weighted by the volume of the efficacy space, which we call $\Omega$,
\begin{equation}\label{eq:marginalization_RDT_V2}
    \langle  T  \rangle =\frac{1}{\Omega} \int_{\Omega} d\pmb{\eta} \, \langle T  \vert \pmb{\eta} \rangle ,
\end{equation}
In the context of the coupled model, the conditioned expectation of $T$ only depends on the radial distance $\eta = \sqrt{\pmb{\eta}\cdot\pmb{\eta}}$, and the volume of the efficacy space reads 
\begin{equation}\label{eq:V_spherical_coordinates}
    \Omega = V(N_T) \left(\eta_\text{max}^{N_T}-\eta_\text{min}^{N_T}\right),
\end{equation}
where $V(N_T)$ is the volume of the hipersphere of $N_T$ dimensions and unit radius. Integrating over angular variables and bearing in mind the Jacobian of the change of coordinates from Cartesian to spherical coordinates, the integral over the efficacy space for the coupled model can be rewritten as
\begin{equation}\label{eq:Jacobian}
    \int_{\Omega} d\pmb{\eta} \, \langle T  \vert \pmb{\eta} \rangle=V(N_T)\,\int_{\eta_\text{min}}^{\eta_\text{max}} d{\eta} \, \eta^{N_T-1} \langle T  \vert {\eta} \rangle.
\end{equation}
Eq.~\eqref{eq:total_average} in the main text is obtained inserting Eqs.~\eqref{eq:V_spherical_coordinates} and~\eqref{eq:Jacobian} in Eq.~\eqref{eq:marginalization_RDT_V2}.

\end{document}